\documentclass[review]{elsarticle}

\usepackage{lineno,hyperref}
\modulolinenumbers[5]

\usepackage{dsfont} 
\usepackage{subfig} 
\usepackage{mathtools} 
\usepackage{graphics} 
\usepackage{multirow}

\usepackage{xcolor}
\usepackage{ulem}

\definecolor{newcolor}{rgb}{.8,.349,.1}

\journal{Journal of \LaTeX\ Templates}

\biboptions{authoryear}

\begin{document}

\begin{frontmatter}

\title{ProstAttention-Net: a deep attention model for prostate cancer segmentation by aggressiveness in MRI scans}

\author[1]{Audrey Duran}
\cortext[cor1]{Corresponding author: 
  Tel.: (+33) 4 72 43 81 48  
  fax: (+33) 4 72 43 85 26;
  }
\author[1]{Gaspard Dussert}
\author[3]{Olivier Rouvière}
\author[3]{Tristan Jaouen}
\author[2]{Pierre-Marc Jodoin}
\author[1]{Carole Lartizien\corref{cor1}}
\ead{carole.lartizien@creatis.insa-lyon.fr}

\address[1]{Univ Lyon, CNRS, Inserm, INSA Lyon, UCBL, CREATIS, UMR5220, U1206, F‐69621, Villeurbanne, France} 
\address[2]{Computer Science Department, University of Sherbrooke, 2500 Boulevard de l'Université, Sherbrooke QC J1K 2R1, Canada}
\address[3]{Department of Urinary and Vascular Imaging, Hospices Civils de Lyon, Hôpital Edouard Herriot, Lyon, France}

\begin{abstract}
Multiparametric magnetic resonance imaging (mp-MRI) has shown excellent results in the detection of prostate cancer (PCa). 
However, characterizing prostate lesions aggressiveness in mp-MRI sequences is impossible in clinical practice, and biopsy remains the reference to determine the Gleason score (GS).
In this work, we propose a novel end-to-end multi-class network that jointly segments the prostate gland and cancer lesions with GS group grading. 
After encoding the information on a latent space, the network is separated in two branches: 1) the first branch performs prostate segmentation 2) the second branch uses this zonal prior as an attention gate for the detection and grading of prostate lesions.
The model was trained and validated with a 5-fold cross-validation on an heterogeneous series of 219 MRI exams acquired on three different scanners prior prostatectomy. 
In the free-response receiver operating characteristics (FROC) analysis for clinically significant lesions (defined as GS~$> 6$) detection, our model achieves 69.0\%~$\pm 14.5$\% sensitivity at 2.9 false positive per patient on the whole prostate and 70.8\%~$\pm 14.4$\% sensitivity at 1.5 false positive when considering the peripheral zone (PZ) only.
Regarding the automatic GS group grading, Cohen's quadratic weighted kappa coefficient ($\kappa$) is $0.418 \pm 0.138$, which is  
the best reported lesion-wise kappa for GS segmentation to our knowledge. 
The model has encouraging generalization capacities with $\kappa=0.120\pm0.092$ on the PROSTATEx-2 public dataset and achieves state-of-the-art performance for the segmentation of the whole prostate gland with a Dice of $0.875\pm0.013$.
Finally, we show that ProstAttention-Net improves performance in comparison to reference segmentation models, including U-Net, DeepLabv3+ and E-Net. The proposed attention mechanism is also shown to outperform Attention U-Net.
\end{abstract}


\end{frontmatter}


\section{Introduction}
\label{section:Introduction}

Prostate cancer (PCa) is the most frequently diagnosed cancer
in men in more than half the countries of the world \citep{sung_global_2021}. With nearly 1.4 million new cases and 375,000 deaths worldwide, it is the second
most frequent cancer and the fifth leading cause of cancer
death among men in 2020. 

Multiparametric magnetic resonance imaging (mp-MRI), which combines T2-weighted imaging with diffusion-weighted and dynamic contrast material–enhanced imaging, has shown excellent results in the detection of high-grade PCa \citep{ahmed_diagnostic_2017, kasivisvanathan_mri-targeted_2018, rouviere_use_2019, van_der_leest_head--head_2019, drost_prostate_2020, klotz_comparison_2021}. As a result, the latest version of the European Association of Urology guidelines now recommends, in case of clinical suspicion of PCa, to perform a prostate mp-MRI prior to any biopsy \citep{mottet_eau-eanm-estro-esur-siog_2021}.
However, characterizing focal prostate lesions in mp-MRI sequences is time demanding and challenging, even for experienced readers, especially when individual MR sequences yield conflicting findings.
Despite the use of the diagnostic criteria from the PI-RADS version 2 scoring system, its inter-reader reproducibility remains moderate at best \citep{richenberg_primacy_2019}. Also, mp-MRI has a low specificity, that could induce unnecessary biopsies \citep{drost_prostate_2020}.

There has been a considerable effort, in the past decade, to develop computer aided detection (CADe) and diagnosis (CADx) systems of PCa as well as prostate segmentation \citep{castillo_t_automated_2020, wildeboer_artificial_2020}. CAD systems for lesion classification from an input region of interest (CADx) have been widely studied, using machine learning \citep{dinh_characterization_2018, ellmann_computer-aided_2020} or deep learning approaches \citep{chen_transfer_2017, le_automated_2017, song_computer-aided_2018,yuan_prostate_2019, zhong_deep_2019,aldoj_semi-automatic_2020}. 

In this paper, we focus our review on the CAD systems performing detection and segmentation of PCa lesions (CADe).
The vast majority of developed CAD models target the detection and segmentation of clinically significant (CS) cancers, i.e. those with a Gleason score $>6$ \citep{ploussard_contemporary_2011}, where the Gleason score (GS) characterizes the cancer aggressiveness \citep{kohl_adversarial_2017, yang_co-trained_2017, wang_automated_2018,schelb_classification_2019, chen_automatic_2020, dai_segmentation_2020,hiremath_test-retest_2020, chiou_harnessing_2021,saha_end-to-end_2021}. The current lesion grading system is based on GS groups, where GS 7 is separated in two groups (GS 3+4 and GS 4+3) and GS 9 and GS 10 are gathered in the same group \citep{epstein_contemporary_2016}.

Despite an important improvement brought by CAD systems that automatically map CS PCa lesions, there is a need to go further by also predicting the degree of PCa aggressiveness that influences patient management \citep{mottet_eau-eanm-estro-esur-siog_2021}.
This task is non-trivial since lesion aggressiveness is not visible on mp-MRI and the classes are highly correlated.
In this work, we propose a novel end-to-end deep architecture called \textit{ProstAttention-Net} that automatically segments the prostate and uses its prediction as an anatomical prior for the detection and grading of prostate lesions. We show that the addition of this attention mechanism improves the CAD performance. 

The main contributions can be summarized as follows:
\begin{itemize}
\item We propose a novel end-to-end architecture based on convolutional neural networks (CNN) with an attention mechanism that performs jointly multi-class segmentation of PCa lesions and prostate segmentation\footnote[1]{Code is available at https://github.com/xxxx};
\item This model is evaluated on a heterogeneous PCa database of 219 patients (1.5T and 3T scanners from 3 manufacturers) with whole-mount histopathology slices of the prostatectomy specimens as ground truth;
\item Performance for the multi-class segmentation task based on Cohen's kappa score at the GS group level is shown to outperform state-of-the-art segmentation architectures;
\item Generalization performance for the multi-class segmentation task on the PROSTATEx-2 challenge database ranks among the best models, without any fine-tuning on PROSTATEx-2 training data.
\end{itemize}

This paper constitutes an extended version of \citet{duran2020prostate}.
In this work, we train our segmentation model on a larger and more heterogeneous dataset, composed of 219 patients (versus 98 patients) acquired on three distinct scanners from different manufacturers. The model is changed with an attention zone that considers the full prostate - peripheral zone (PZ) and transition zone (TZ) - while we focused on PZ and PZ lesions in our previous work. The number of classes is shifted from 7 to 6 classes by gathering GS~$8$ and GS~$\ge9$ in the same group to compensate the lower number of lesions in those groups. In terms of performance, lesions detection for both prostate zones (PZ and TZ) are analysed separately. Generalization performance on an external dataset is assessed by testing the model on the PROSTATEx-2 challenge dataset. 
We also include comparison with state-of-the-art segmentation algorithms including U-Net, DeepLabv3+, E-Net and Attention U-Net. Finally, performance metrics are extended with the addition of the confusion matrix at the lesion-level and its associated lesion-wise Cohen's kappa score. 

\vspace{-0.2cm}
\section{Related works}
\vspace{-0.2cm}
\label{section:relatedworks}

\paragraph{Segmentation networks with attention mechanisms}
Attention mechanisms are commonly used in computer vision for classification and segmentation problems \citep{chaudhari_attentive_2020}. They are aiming at emphasizing salient features while dimming the useless ones, mimicking human visual attention. 
Some attention mechanisms have shown improvement of segmentation performance in medical imaging problems. The soft attention gates proposed in \citet{schlemper_attention_2019} lead to a 2-3\% Dice improvement and a better recall in comparison to U-Net baseline for segmentation of a 150 3D-CT abdominal image dataset. 
The adaptation of the "squeeze \& excitation" (SE) modules \citep{Hu_2018_CVPR} for 3 imaging segmentation problems by \citet{roy_recalibrating_2019} increases the Dice score by 4-9\% in the case of U-Net. More specifically for prostate application, these modules were also employed by \citet{rundo_use-net_2019} for prostate zonal segmentation of heterogeneous MRI datasets, and lead to a 1.4-2.9\% Dice increase in comparison to U-Net baseline for PZ segmentation when evaluating their multi-source model on different datasets. In \citet{zhang_bi-attention_2019}, the combined channel-attention (inspired by SE modules) and position-attention layers reached a Dice of 1.8\% higher than their baseline for the segmentation of prostate lesions on T2w MR. \citet{de_vente_deep_2020} tested different strategies to focus the attention on each prostate zone (PZ and TZ) separately. Performance of their model slightly increased when merging the zonal feature maps, corresponding to probabilistic segmentation maps derived from a separated network, before
the final convolution (with a voxel-based kappa score increasing from $0.391\pm0.062$ to
$0.400\pm 0.064$) but were not statistically better than when omitting zonal information.
\citet{saha_end-to-end_2021} could increase the sensitivity at 1 false positive per patient of 4.34\% by the addition of the previously mentioned SE and grid-attention gates attention mechanisms in a 3D UNet++ \citep{zhou_unet_2020} backbone architecture for binary PCa segmentation. 
But the most important performance gain was due to the addition of a probabilistic anatomical prior as input channel of the feature encoding branch of their model. This prior which captures the spatial prevalence and zonal distinction of CS PCa led to an extra 4.10\% of detected lesions.

\paragraph{2-steps workflow for PCa detection} Recent studies demonstrate good performance of architectures combining two cascaded networks: one segmenting or performing a crop around the prostate, followed by a binary PCa lesion segmentation model. In \citet{yang_co-trained_2017}, a regression CNN is used to crop a square region encompassing the entire prostate region. Then, each pair of beforehand aligned ADC and T2w squares is input into their proposed co-trained CNNs. This work was extended in \citet{wang_automated_2018}, where they adapted the workflow into an end-to-end trainable deep neural network, composed of two sub-networks:
the first one performs both prostate detection and ADC-T2w registration, and the second one is a dual-path multimodal CNN producing CS/non CS classification score and CS-class response map. The study by \citet{hosseinzadeh:MIDLAbstract2019a} uses a 
preparatory network (anisotropic 3DUNet) to produce deterministic or probabilistic zonal segmentation. Those maps are then fused in a second CNN outputting a CS PCa probability map. However, a 2-steps workflow implies additional weights and a heavier training workflow so that the most recent architectures are based on one-step segmentation networks.

\paragraph{PCa classification by aggressiveness}
A few networks go beyond the binary classification of CS PCa and predict cancer aggressiveness.
In 2017, the PROSTATEx-2 challenge ~\citep{prostatex2} was proposed to classify regions of interest (suspicious lesions) by Gleason score. 
The winners \citep{abraham_computer-aided_2018} proposed a method where high-level features are extracted from hand-crafted texture features using a deep network of stacked sparse autoencoders and classified using a softmax classifier. It achieves a quadratic-weighted kappa score (see definition in Section \ref{subsection:perfmetrics}) of 0.2326 with 3-fold cross-validation using training data of the challenge and 0.2772 when evaluated on the 70 challenge test lesions. In \citet{abraham_automated_2019}, the same team used a VGG-16 CNN followed by an ordinal class classifier for PCa classification. They achieved a weighted kappa score of 0.4727 in a Leave-One-Patient-Out (LOPO) cross-validation strategy on PROSTATEx-2 train set, which is the highest reported kappa on this dataset.
More recently, \citet{chaddad_deep_2020} proposed a new classification framework that uses deep entropy features (DEFs) derived from CNNs to predict the Gleason score of PCa lesions. The combined DEFs of all CNNs are used as input to a random forest classifier that can discern GS of 6, 3+4, 4+3, 8 and $\ge$~9 with an AUC (\%) of 80.08, 85.77, 97.30, 98.20, and 86.51 respectively when trained and tested on PROSTATEx-2 train dataset.

\paragraph{PCa detection or segmentation by aggressiveness}
The detection or segmentation of PCa by aggressiveness has been addressed by very few studies.
\citet{cao_joint_2019} proposed a multi-class CNN built on DeepLab-ResNet101 architecture, FocalNet. The model is trained and validated on a private 417 patients dataset (728 lesions and 1750 slices) with ordinal encoding, to exploit the hierarchy among GS classes. It produces a five-channel prediction of cancer aggressiveness. From this pixel-level output, lesion localization points are extracted, thus performing lesion detection and not segmentation. Based on these detection points, FocalNet reaches 87.9\% sensitivity for CS lesions at 1 false positive per patient when the evaluation is performed on 2D slices containing at least one reported lesion (slices without any reported lesion are discarded).
Considering the GS prediction, FocalNet reaches AUC of $0.81\pm0.01$, $0.79\pm0.01$, 0.67$\pm$0.04, and 0.57$\pm$0.02 for the binary classification tasks GS~$\ge$7 vs. GS~$<$7, GS~$\ge$4+3 vs. GS~$\le$3+4, GS~$\ge$8 vs. GS~$<$8 and GS~$\ge$9 vs. GS~$<$9 respectively when considering detected lesions only.

\citet{de_vente_deep_2020} proposed a model for lesion segmentation by GS based on a multi-label regression U-Net associated to an ordinal encoding as in \citet{cao_joint_2019}. They adapted the PROSTATEx-2 classification challenge dataset for segmentation task by delineating ground truth lesions from the lesion centroid coordinates provided in the challenge using in-house software. 
They also studied the incorporation of zonal information as attention mechanism with different strategies but did not show any benefit. 
They reached a lesion-wise weighted kappa of $0.172 \pm 0.169$ with 5-fold cross-validation on PROSTATEx-2 train set and $0.13 \pm 0.27$ on the test set.

\paragraph{Performance evaluation of PCa CAD models : PCa databases, quality of the ground truth annotations and generalization}
There are very few public databases dedicated to the task of PCa segmentation. Most of it include a small amount of patients (e.g. 42 for I2CVB \citep{lemaitre_computer-aided_2015}, 28 for Prostate Fused-MRI-Pathology \citep{madabhushi_fused_2016}), provide binary annotations (CS PCa versus normal tissue) and are based on ground truth obtained with biopsy samples.
The PROSTATEx-2 challenge dataset (see description in Section \ref{subsubsection:Performance_Px2_dataset_Mat}) includes a significant number of patient exams (99 train and 63 test patients) and biopsy-based GS group grading, but it does not provide lesion contours since it was designed for a classification challenge.
As a result, most of the PCa segmentation models are trained and validated on private datasets, often homogeneous, without test on external data from other institutions and acquired on scanners of different manufacturers \citep{tsehay_biopsy-guided_2017, song_computer-aided_2018, wang_automated_2018,cao_joint_2019,schelb_classification_2019,de_vente_deep_2020}. This does not guarantee good generalization on test data sampled from different distributions. 
Some papers evaluate their CAD model on public datasets \citep{wang_automated_2018, de_vente_deep_2020, netzer_fully_2021}, but as far we know, none of them without using it in the training process, which impairs generalization assessment.

Another limitation of the current evaluation of PCa CAD models is the biopsy-based ground truth estimation. As emphasized in \citet{cao_joint_2019}, the biopsy-based datasets are biased since biopsy cores are mostly based on MRI-positive findings. As a consequence, undetected lesions on the MRI are often missed. This may result in an overestimation of the performance of the CAD systems. 
In addition, the biopsy sample GS estimation might not be representative of the real lesion GS. Indeed, \citet{epstein_upgrading_2012} reported that more than one-third of the biopsy cases with GS~$\le$6 were upgraded to GS~$\ge$7, and one-fourth of GS 3+4 in biopsy were downgraded after checking with whole-mount histopathology. This is why whole-mount histopathology is the gold standard (but hard to acquire) ground truth for lesion detection and characterization. Also note that a prostatectomy-based dataset contains more PCa lesions than a biopsy-based dataset including the same number of patients. The prevalence of PCa is indeed higher in a patient population referred for prostatectomy than in one referred for biopsy. As such, our private  219-patient dataset contains 234 CS PCa compared to 97 CS PCa for \cite{saha_end-to-end_2021} biopsy-based dataset, that includes 296 patients.

\section{Method}
We propose an end-to-end network that performs multi-class segmentation and exploit the spatial context for PCa detection.
We first present the attention model architecture, with the used backbone CNN architecture and the global loss, followed by the implementation details. Finally, the post-processing of the output labeled maps is explained.

\subsection{ProstAttention-Net for PCa and prostate segmentation}

\begin{figure*}[htbp]
\centering
\caption{\label{fig:attunet}Our proposed ProstAttention-Net model. On the left is the encoder, on the top right is the prostate decoder and on the bottom right is the lesion decoder. The encircled X stands for the attention operation. The pink region superimposed on the T2W MRI on the upper right decoding branch corresponds to the output prostate gland segmented by the network. The pink and green regions superimposed on the T2w MRI in the lower right decoding branch correspond respectively to the prostate and a GS~3+4 lesion detected by the network.}
\includegraphics[width=\linewidth]{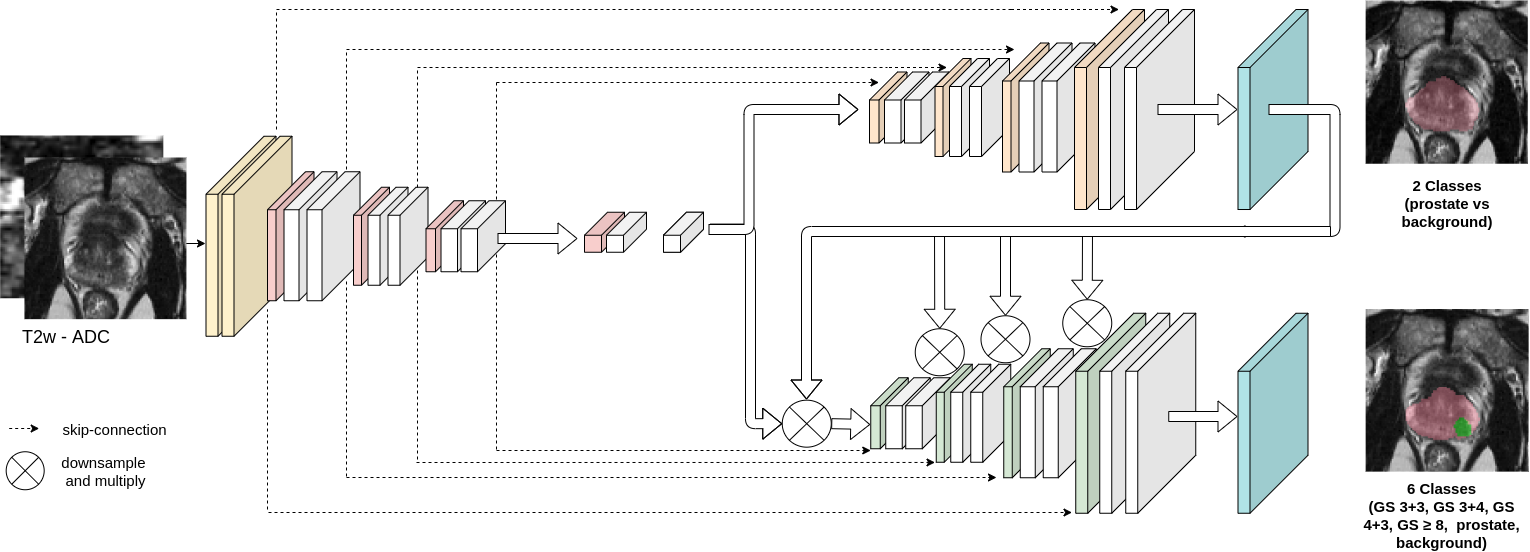}  
\end{figure*}

As depicted on \autoref{fig:attunet}, our ProstAttention-Net model is an end-to-end multi-class deep network that jointly performs two tasks: 1) the segmentation of the prostate gland and 2) the detection, segmentation and GS group grading of prostate lesions. 

The encoder of our network first encodes the information from multichannel T2w and ADC input images into a latent space.
This latent representation is then connected to two decoding branches: 
\begin{itemize}
    \item the first one performs a binary segmentation of the prostate gland
    \item the second one uses the predicted prostate segmentation as a prior through an attention mechanism for the detection and grading of prostate lesions. 
\end{itemize}

The output probabilistic map of the prostate decoder serves as a soft attention map to the lesion decoder branch. This self trained attention map is first downsampled to the resolution of each block of the lesion branch and multiplied to the input feature maps of this block. This attention operation is a channel-wise Hadamard product between the resampled output of the prostate branch and the feature maps of the second branch.

This is somewhat inspired but different from the Attention U-Net \citep{oktay_attention_2018, schlemper_attention_2019}, as our attention map comes from the resampling of the prostate prediction instead of the preceding convolutional block. The idea here is to enforce the tumor prediction to be within the prostate by shutting down the neurons located outside the prostate.
This approach can be compared to the probabilistic anatomical prior used in \cite{saha_end-to-end_2021}.  
As reported in Section \ref{section:relatedworks}, \cite{saha_end-to-end_2021} indeed modeled the spatial distributions of the prostate zone and malignant regions estimated on their dataset and showed that this spatial prior used as a extra channel of their feature encoding network allowed significant performance gain.
This study confirms empirical evidences showing that the anatomical context helps the network in the learning phase. In addition, the fact that the system outlines the prostate region increases the credibility of the network and thus enforces the confidence radiologists may have in the system.

\subsection{Backbone CNN architecture}
The backbone CNN architecture of our attention model is based on U-Net \citep{ronneberger_u-net_2015}, with batch normalization \citep{Ioffe2015BatchNA} layers to reduce over-fitting.
The encoder part of our model contains five blocks, each composed of two convolutional layers with kernel size $3\times3$, followed by a Leaky ReLu  \citep{Maas2013RectifierNI} and MaxPool layers.
The prostate decoder branch follows the same architecture but with transposed convolutions to increase the feature maps resolution. Its output has two channels corresponding to the prostate and background classes.
The lesion decoder has an architecture similar to the prostate decoder branch except that it produces 6-channels segmentation maps, corresponding to the background, the overall prostate area, GS 6, GS 3+4, GS 4+3 and GS~$\ge8$ lesions with class labels $c$ ranging from 1 to 6, respectively.

\subsection{Global loss}
The global loss of the ProstAttention-Net is defined as 
\begin{equation}
 L = \lambda_1.L_{prostate} + \lambda_2.L_{lesion}
 \label{eq:loss}
\end{equation}
where $L_{prostate}$ and $L_{lesion}$ are the losses corresponding to the prostate and lesion segmentation task and $\lambda_1$ and $\lambda_2$ are weights to balance between both losses whose value can be varied during training. The $L_{prostate}$ and $L_{lesion}$ loss functions were defined as the sum of the crossentropy and Dice losses. To deal with class imbalance, each loss term was weighted by a class-specific weight $w_c$. These losses can be detailed as:
\begin{equation}
L_{prostate} = 1 - 2 \frac{\sum_{c=1}^{2} w_c \sum_{i=1}^{N}y_{ci}p_{ci}}{\sum_{c=1}^{2} w_c \sum_{i=1}^{N}y_{ci} +  p_{ci}}- \frac{1}{N}\sum_{i=1}^{N}\sum_{c=1}^{2}\mathds{1}_{y_i \in C_c}w_c\log p_{ci}
\label{eq:lossprostate}
\end{equation}
\begin{equation}
L_{lesion} = 1 - 2 \frac{\sum_{c=1}^{6} w_c \sum_{i=1}^{N}y_{ci}p_{ci}}{\sum_{c=1}^{6} w_c \sum_{i=1}^{N}y_{ci} + p_{ci}} -\frac{1}{N}\sum_{i=1}^{N}\sum_{c=1}^{6}\mathds{1}_{y_i \in C_c}w_c \log p_{ci}
\label{eq:losslesion}
\end{equation}

where $w_c$ is the class-specific weight, $p_{ci}$ the probability predicted by the model for the observation $i$ to belong to class $c$ and $y_{ci}$ is equal to 1 if pixel $i$ belongs to class $c$, and 0 otherwise (the ground truth label for pixel $i$).

\subsection{Implementation details}

Input data was preprocessed as described in Section \ref{subsubsection:data_preprocessing}.

Concerning loss weights defined in \autoref{eq:loss}, $\lambda_1$ was set to 1 and $\lambda_2$ to 0 during the first 20 epochs in order to focus the training on the prostate segmentation task. Then, $\lambda_2$ was changed to 1 to allow an equal contribution of the $L_{prostate}$ and $L_{lesion}$ loss terms.
Class-specific weights $w_c$ were set to 0.002 for background, 0.14 for prostate and 0.1715 for each lesion class based on the class pixels frequency estimated from our clinical dataset (see Section \ref{subsection:dataset}).

The whole network was trained end-to-end using Adam \citep{2014arXiv1412.6980K} and a L2 weight regularization with $\gamma= 10^{-4}$. The initial learning rate was set to $10^{-3}$ with a 0.5 decay after 25 epochs without validation loss improvement. At the end, we kept the weights associated to the highest validation Dice (as in \autoref{eq:losslesion} but without the background class and the weight factors $w_c$). These hyperparameters ($\lambda_1$, $\lambda_2$, $\gamma$, the learning rate, the epoch of inclusion of the second branch, $w_c$ etc.) were tuned with random grid search.
The pipeline was implemented in python with the Keras-Tensorflow library \citep{chollet2015keras,tensorflow2015-whitepaper}.

\subsection{Post-processing of the output labeled maps}
\label{subsection:postprocessing}

The decoder branch of ProstAttention-Net outputs a label map where each voxel is assigned the class label corresponding to the maximum probability value of the softmax output layer. A 3D labeled volume per patient is then reconstructed by stacking all 2D labeled transverse slices of this patient. This constitutes the \textit{raw 3D labeled maps}. 
\textit{3D lesion maps }can then be estimated from these raw labeled volumes by identifying the connected components. Depending on the clinician need, two types of cluster maps may be generated : 
\begin{itemize}
    \item the \textit{GS lesion maps} are multi-class lesion maps where one lesion is a cluster of neighboring voxels with the same GS.  This is achieved by applying the clustering process on each GS label maps extracted from the \textit{raw 3D labeled maps}, that is clustering independently all voxels of one particular GS class.
    \item the \textit{CS lesion maps} are binary lesion maps corresponding to CS cancer only. They are computed by first thresholding the raw 3D labels maps to consider only voxels with class label corresponding to GS~$>6$ and then applying the clustering process on these binary CS lesion masks.
\end{itemize}

A simple post-processing of these cluster maps consists in discarding clusters smaller than a certain pre-determined volume.  Other post-processing strategies may be applied based on the performance metrics estimated on the trained model (refer to Section \ref{subsection:perfmetrics}) and the clinician's need. This include, for example, setting a particular point on the FROC curves, corresponding to a fixed authorized mean false positive rate and considering the corresponding threshold to remove all detected clusters with lower lesional probability.

In the following, we considered a 63-connectivity rule for the clustering process and removed all clusters whose volume is below 45 mm$^3$ (15 voxels). This value was derived from the distribution of the cancer lesion sizes of our private dataset, as detailed in Section \ref{subsubsection:dataset_lesionannotation}. 
\section{Experimental settings}

\subsection{Dataset description and preprocessing}
\label{subsection:dataset}
This Section presents the overall dataset and annotation process. Note that a more detailed protocol (preparation of the prostatectomy specimen, histopathological analysis, MR image analysis, etc.) is described in \citet{bratan_influence_2013}.

\subsubsection{MRI Dataset}
The dataset consists in a series of mp-MRI exams of 219 patients, acquired in clinical practice in two departments of radiology at our partner clinical center since September 2008. It was declared to the appropriate national administrative authorities (\textit{Comité de Protection des Personnes, reference L 09-04} and \textit{Commission Nationale de l’Informatique et des Libertés, treatment n$^{\circ}$ 08-06}) and patients gave written informed consent for researchers to use their MR imaging and pathologic data.
All patients were scheduled to undergo radical prostatectomy, leading to a majority of patients with CS PCa (183/219) and a high number of lesions in the dataset (cf. Section~\ref{subsubsection:dataset_lesionannotation}).

Imaging was performed on three different scanners from different constructors: 67 exams were acquired on a 1.5 Tesla Siemens scanner (Symphony; Siemens, Erlangen, Germany), 126 on a 3 Tesla GE scanner (Discovery; General Electric, Milwaukee, USA) and 26 on a 3 Tesla Philips scanner (Ingenia; Philips Healthcare, Amsterdam, Netherlands).
\begin{table*}[!t]
\caption{\label{tab:paramsMRI}Parameters for prostate imaging on the different scanners constituting our private dataset}
\centering
\resizebox{\textwidth}{!}{
\begin{tabular}{ccccccccc}
\hline 
Scanner & Field & Sequence & $T_R$  & $T_E$ & FOV & Matrix & Voxel dimension & Max b-value\\ &strength& &(ms)&(ms)&(mm)&(voxels)&(mm)&(s/mm$^2$)\\
\hline
Siemens Symphony & 1.5T & T2w & 7750 & 109 & $200\times200$ & $256\times256$ & $.78\times.78\times3$& -\\
Siemens Symphony & 1.5T & ADC& 4800 & 90 & $300\times206$ & $128\times88$ & $2.34\times2.34\times3$& 600\\
GE Discovery & 3T & T2w & 5000 & 104 & $220\times220$ & $512\times512$ & $.43\times.43\times3$ & -\\
GE Discovery & 3T & ADC & 5000 & 90 & $380\times380$ & $256\times256$ & $1.48\times1.48\times3$& 2000 \\
Philips Ingenia & 3T & T2w & $\sim$ 5500 & 100 & $180\times180$ & $336\times336$ & $.54\times.54\times3$ & -\\
Philips Ingenia & 3T & ADC & $\sim$ 4800 & $\sim$ 81 & $350\times350$ & $288\times288$ & $1.22\times1.22\times3$ & 2000
\\ \hline
\end{tabular}
}
\end{table*}
Data includes axial T2 weighted (T2w) and apparent diffusion coefficient (ADC) maps computed from the diffusion weighted imaging (DWI) sequence. The ADC maps were obtained by using linear least squares curve fitting of pixels (in log scale) in the two or five available diffusion-weighted images against their corresponding b-values. Parameters of the sequences are reported in \autoref{tab:paramsMRI}, as it varied as per the standard of care at the time of the examination.

\subsubsection{MRI preprocessing}
\label{subsubsection:data_preprocessing}
T2w and ADC images were spatially resampled to a common
axial in-plane resolution of $1\times1\times3$ mm voxel size and automatically cropped to a $96\times96$ pixels region on the image's center. Intensity was linearly normalized into [0, 1] by channel. 
Negligible patient motion across the different sequences was observed. Thus, no additional registration techniques were applied, in agreement with clinical recommendations \citep{epstein_2014_2016} and recent studies \citep{cao_joint_2019, saha_end-to-end_2021}. 

\subsubsection{Prostate gland and PCa lesions annotation}
\label{subsubsection:dataset_lesionannotation}
\begin{figure*}[!t]
 \caption{\label{fig:anapath}Prostate MRI annotation. From left to right: T2w MR image, ADC maps, final prostate annotation overlapped on T2w and histology slice.}
  {\includegraphics[width=\linewidth]{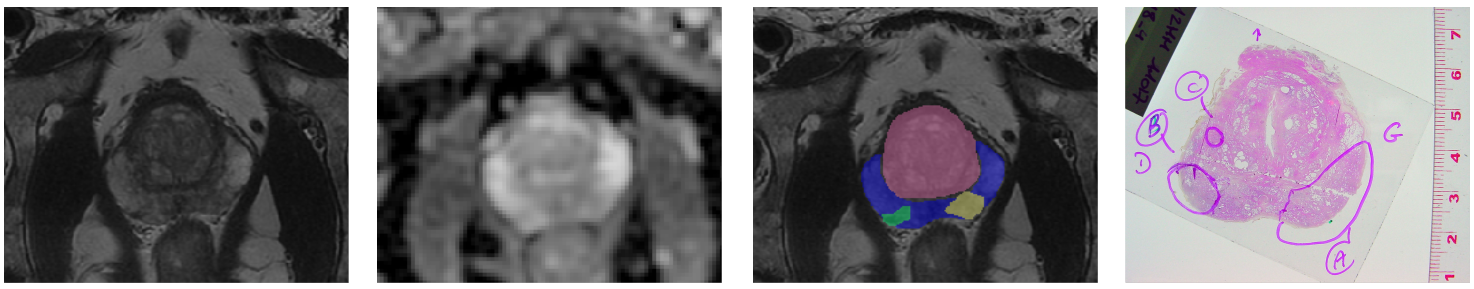}}
\end{figure*}

The 219 MRI exams of our dataset were reviewed by two uroradiologists (11 and 1 year of experience in prostate imaging at the database creation) who outlined  prostate focal lesions that showed abnormal signal on images (low signal intensity on T2w and/or low intensity on ADC maps and/or early enhancement on DCE images).
They were blinded to the clinical and pathologic data.
The prostatectomy specimens were analyzed \textit{a posteriori} by an anatomopathologist (10 years of experience when the database was created) according to international guidelines \citep{samaratunga_international_2011} thus providing the histological ground truth as seen on \autoref{fig:anapath}.
After correlation with the whole-mount specimens, the uroradiologists reported in consensus 342 lesions, where 293 belong to the PZ. Note that 71 and 92 lesions among the 342 reported lesions were missed by the 11 years and 1 year of experience uroradiologist respectively during their initial reading.
Lesions that were not visible on mp-MRI exams \textit{a posteriori} were not reported. In total, 67 lesions were visible in the whole-mount specimens but could not be identified on mp-MRI during the consensus reading, mainly of GS 6 (corresponding to 35\% of the GS 6 lesions identified on histopathology specimen).
Lesion volumes range between 9 and 9918 voxels, with a mean size of $627\pm1026$ voxels and a median size of 306 voxels.
As a post-processing, lesions smaller than 15 voxels (45 mm$^3$) were discarded (as explained Section~\ref{subsection:postprocessing}), thus leading to 338 lesions in the dataset including 289 in the PZ. Their distribution according to the GS group is detailed in \autoref{tab:lesionsclass}.
In addition to lesions annotation, the radiologists also contoured the TZ and the PZ on each volume.
All patients acquired on Siemens Symphony, GE Discovery or Philips Ingenia scanners were included without consideration of the MRI quality. Patients with a prosthetic or with blood on the MRI were excluded.

\begin{table}[!t]
\caption{\label{tab:lesionsclass}Lesions distribution by Gleason Score for all patients of our private dataset, corresponding to 67 Siemens Symphony, 126 GE Discovery and 26 Philips Ingenia acquisitions.}
\centering
\begin{tabular}{cccccc}
\hline
 & \bfseries GS 3+3 &  \bfseries GS 3+4 & \bfseries GS 4+3  & \bfseries \boldmath GS~$\ge8$ & \bfseries Total\\ 
PZ & 83 & 111 & 52 & 43 & 289\\
TZ & 21 & 15 & 4 & 9 & 49 \\ 
Total & 104 & 126 & 56 & 52 & 338 \\
\hline
\end{tabular}
\end{table}

\subsection{Experiments}
\subsubsection{Performance analysis of ProstAttention-Net}
\label{subsection:base_expe_prostatttention}

ProstAttention-Net was trained and tested with the 219 patients database including 338 lesions described in Section~\ref{subsection:dataset} following a 5-fold cross-validation strategy. 
We repeated the cross-validation experiment with the best combination of hyperparameters and 
same data split to obtain 4 replicates. The cross-validation experiment producing the highest Cohen's kappa score (see Section~\ref{subsection:perfmetrics}) on the validation set was selected. Each fold contained 43 or 44 patients (i.e. around 1000 slices) and was balanced regarding lesions classes and the type of MRI scanner. A preliminary study (cf. \ref{appendix:multisource}) confirmed the benefit of multi-source learning with data from all scanners instead of single-source models. Training and evaluation were conducted on the whole patient 3D volumes, each constituted of 24 transverse slices including slices without visible prostate or without lesion annotation, leading to 4000 training slices. For comparison, the training dataset in \cite{cao_joint_2019} consists of 1400 slices retrieved from 417 patients, since only the slices with at least one lesion are considered. No prostate mask was applied to the input bi-parametric MRI images or output prediction maps, both during training or evaluation phases if not specified.

Performance was evaluated first in a binary setting to report the detection of CS lesions and then at the Gleason score level. Impact of the lesion localization (either on the PZ or TZ) was also assessed by applying the zonal masks drawn by the radiologists for each patient. Finally, in order to compare with our previous work \citep{duran2020prostate}, we included the following experiments : 
\begin{itemize}
    \item \textbf{Attention on the PZ} : ProstAttention-Net was trained with an attention model focused on the peripheral zone only to focus on the detection of PZ lesions (corresponding to the vast majority of prostate lesions) as in \citet{duran2020prostate}. 
    \item \textbf{Evaluation on a 98 patients dataset}, corresponding to the subpart of our private dataset used in \citet{duran2020prostate}.
\end{itemize}

We considered different detection and segmentation metrics whose description is provided in Section \ref{subsection:perfmetrics}.

\subsubsection{Comparison with state-of-the-art segmentation networks}
To gauge the impact of our proposed architecture, we report results from 4 other state-of-the-art CNN segmentation models. The first one is the U-Net at the core of our architecture without the attention mechanism.
The second model is E-Net~\citep{paszke_enet_2016}, a CNN for low-latency tasks with an excellent trade-off between accuracy and inference time. The third model is DeepLabv3+ with XceptionNet as a backbone \citep{Chen_2018_ECCV}. DeepLabv3+ is a reference model often used by the computer vision community, that uses spatial pyramid pooling and atrous separable convolutions to capture multiscale characteristics. We also tested DeepLab-ResNet101 as a backbone, the architecture used in \cite{cao_joint_2019}, but got lower performance than with the XceptionNet backbone.
Finally, we also include comparison with Attention U-Net \citep{schlemper_attention_2019}, a U-Net model with attention gates introduced in the skipped connections as reported in Section \ref{section:relatedworks}. Attention gates allow the network learning to suppress irrelevant regions in an input image while highlighting salient features useful for a specific task. 

All model's output have 6-channels, similar to the lesion branch of ProstAttention-Net.
To enable a fair comparison, all steps of the experimental protocol were performed the same way as for ProstAttention-Net, including data pre-processing, hyperparameters search and selection of the best model for each architecture based on 4 replicates of a 5-fold cross validation analysis.
The learning rate was set to  $10^{-3}$ for U-Net and DeepLabv3+, $3.87\times10^{-3}$ for ENet and $1.82\times10^{-2}$ for Attention U-Net. Concerning the regularization weight, the optimal values were $10^{-4}$ for U-Net, DeepLabv3+ and ENet and $1.95\times10^{-5}$ for Attention U-Net.

\subsubsection{Performance on the PROSTATEx-2 dataset}
\label{subsubsection:Performance_Px2_dataset_Mat}
In order to gauge the ability of our method to generalize on images acquired on a different clinical environment, we tested our approach on the PROSTATEx-2 challenge dataset~\citep{prostatex2}. This dataset is composed of a training set of 99 patients with 112 lesions (50 in PZ and 62 in TZ - see \autoref{tab:Px2lesionsclass}) and of a test set containing 63 patients and 70 lesions.
These data were acquired on 3T MAGNETOM Trio and Skyra (Siemens Medical Systems) scanners with different imaging parameters, thus constituting two different sources.  

Details are available in \autoref{tab:paramsMRI_PROSTATEx-2}. 
The ground truth consists of the coordinates of each lesion's center, with its associated Gleason score.
Performance of ProstAttention-Net was estimated on the 99 train patients of PROSTATEx-2, since the challenge submission web page is closed so that we could not estimate performance of our network on the 63 test patients.
\begin{figure*}[!t]
 \caption{\label{fig:diffsources} T2w, ADC and ground truth for a GS 3+4 lesion in the different sources. From left to right: Siemens Symphony, GE Discovery, Philips Ingenia and PROSTATEx-2 (Siemens Skyra) images.}
  {\includegraphics[trim=0 0 150 12,clip, width=\linewidth]{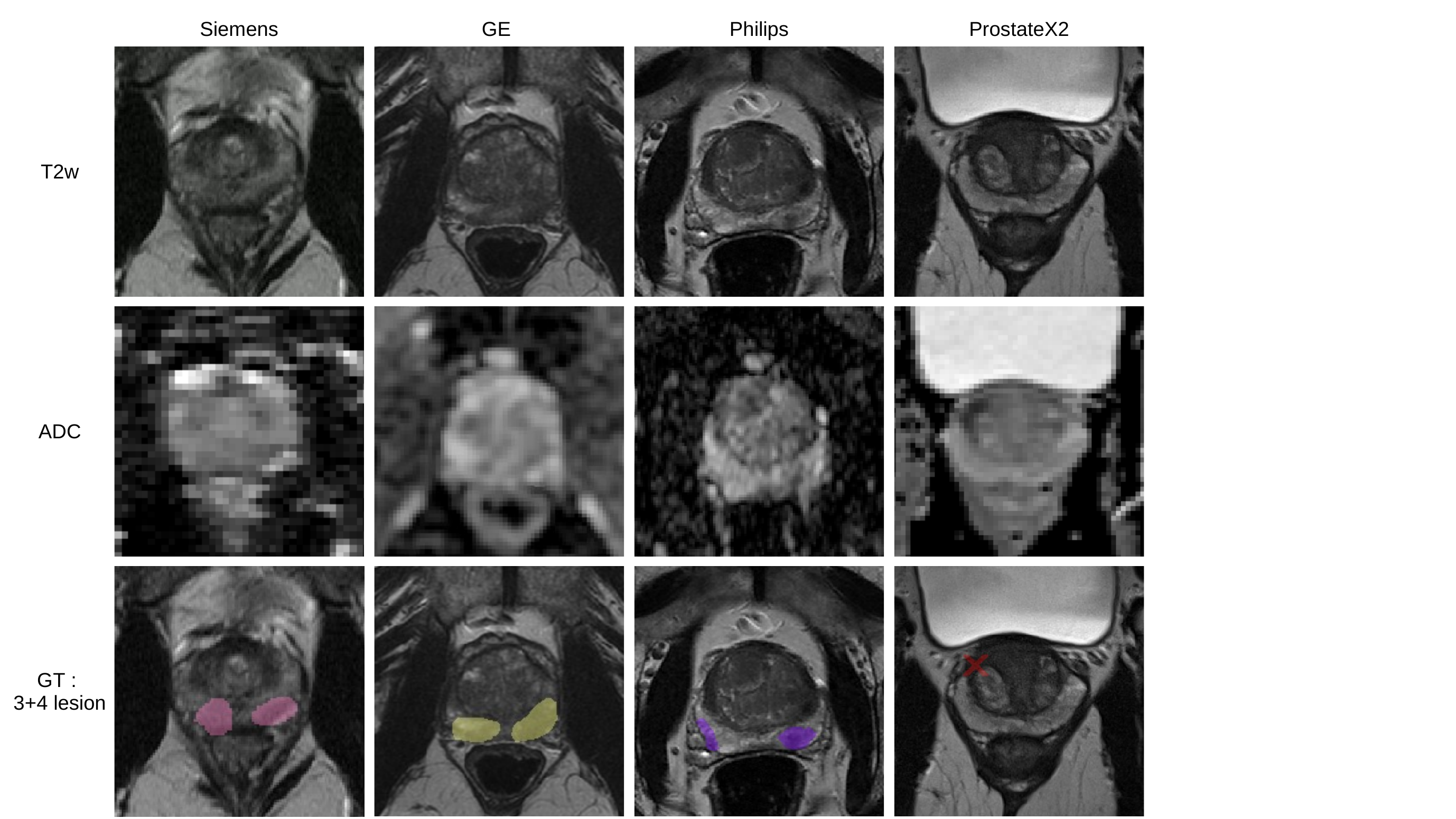}}  
\end{figure*}
From the 5 ProstAttention-Net models obtained by cross-validation with our 219 patients dataset (see Section \ref{subsection:base_expe_prostatttention}), the model with the highest lesion-wise Cohen's kappa score (see definition in part \ref{subsection:perfmetrics}) on the validation set was used to predict classes of PROSTATEx-2 train set lesions. This follows \citet{de_vente_deep_2020} methodology.

\subsection{Performance metrics}
\label{subsection:perfmetrics}

Performance evaluation of all considered architectures was conducted based on standard quantitative detection and segmentation metrics. This includes indices derived from FROC analysis or from confusion matrices, such as Cohen's kappa score for the detection task, or Dice index for the prostate segmentation task.

\paragraph{FROC for lesion detection} Lesion detection performance was evaluated through free-response receiver operating characteristics (FROC) analysis based on the \textit{lesion maps} whose computation is explained in Section \ref{subsection:postprocessing}. FROC curves report detection sensitivity at the lesion level as a function of the mean number of false positive lesion detections per patient.\\
Each detected lesion was assigned a \textit{lesion probability score} corresponding to the average of the voxel probability values in the cluster. FROC curve was then plotted by varying a threshold on this lesion probability score. For each value of this threshold, a lesion was considered a \textit{true positive} (TP) when at least 10\% of its volume intersected a true lesion and if its lesion probability score exceeded the threshold. If it did not intersect a true lesion, it was considered as a \textit{false positive} (FP). We chose this 10\% cut-off value to account for the inter-expert variability (due to indistinct lesion margins and small size of PCa lesions) and to align with other studies \citep{cao_joint_2019,saha_end-to-end_2021}.
Note that thanks to the histopathology-based ground truth, the definition of true or false detection is more accurate than the studies with a biopsy-based ground truth.

Two different FROC analyses were performed :
\begin{itemize}
    \item the first one evaluates the performance of the models to discriminate CS lesions (GS~$>6$). It is computed on the \textit{CS lesion maps} described in Section \ref{subsection:postprocessing}, after removal of the smallest clusters.
    \item the second one evaluates the model ability to discriminate lesions of each GS group. It is computed on the \textit{GS lesion maps} described in Section \ref{subsection:postprocessing}. For the latter, if a lesion was detected at the correct localization but not assigned the correct GS group, it was both considered as a false negative for its ground truth class and a false positive for the predicted class. Note that this evaluation is very pessimistic. 
\end{itemize}

In order to compare the different FROC values, one-sided Wilcoxon signed-rank statistical test was applied to compare the obtained mean sensitivity of each cross-validation fold at two given FP rates (1 and 1.5 FP) for the four GS groups.

\paragraph{Confusion matrix and Cohen's kappa score for GS prediction}
To evaluate the agreement between the ground truth and the prediction, confusion matrices were computed for the following four classes: GS 6, GS 3+4, GS 4+3 and GS~$\ge8$. 
From the \textit{GS lesion maps}, each true positive ground truth lesion was assigned a prediction class. If the ground truth lesion intersected several predicted lesions, the predicted lesion with the highest Dice with the ground truth was considered. 
In order to compare with \citet{de_vente_deep_2020}, we also computed a confusion matrix that includes false negative detections as well. In this configuration, lesions that were missed by the deep segmentation model were considered as predicted in the GS 6 class.
Then, quadratic weighted Cohen's kappa coefficient of agreement was computed from this 4-classes confusion-matrix, as proposed in the PROSTATEx-2 challenge.
Weighted Cohen's kappa takes into account the class-distance between the ground truth and the prediction and allows disagreements to be differently weighted, which is useful when classes are ordered. As an example, a GS 6 lesion wrongly predicted as GS~$\ge8$ (high disagreement) will be more penalized by the weighted kappa metric than a GS 6 lesion wrongly predicted as GS~3+4 (lower disagreement). The kappa metric ranges from -1.0 to 1.0 : the maximum value means complete agreement; zero means chance agreement; and -1.0 means complete disagreement.
For all experiments, the reported kappa value is the average coefficient of the 5 kappa scores, each derived from the confusion matrix estimated on each validation fold, along with the corresponding standard deviation. For some models, we also report total confusion matrices, obtained by adding up the 5 confusion matrices computed on each validation fold, as well as their corresponding kappa scores. 

\paragraph{Evaluation on the PROSTATEx-2 challenge dataset}

As lesion contours are not available for PROSTATEx-2 train and test datasets, we followed the evaluation method proposed by \citet{de_vente_deep_2020}. For each reference lesion center (whose coordinates are provided by the PROSTATEx-2 challenge organizers), a true positive was reported if these coordinates corresponded to a voxel detected as CS in the \textit{CS lesion maps}. The assigned class was then set to the most represented GS class of the cluster. If the reference lesion center did not intersect any detected lesion in the \textit{CS lesion maps}, then it was reported as GS 6 lesion. Performance analysis was then conducted based on Cohen's kappa scores derived from the confusion matrix with 1000 iterations bootstrapping.

\section{Results}

\subsection{Performance of ProstAttention-Net}

\subsubsection{Prostate segmentation}
The prostate segmentation outputted by the first upper branch is evaluated with the average Dice obtained on each validation subfold. ProstAttention-Net obtains results close to state of the art with a Dice of $0.875 \pm 0.013$.

\subsubsection{Detection of clinically significant lesions}
\label{subsubsection:results_cslesions}
\begin{figure}[tp]
\centering
\caption{\label{fig:froc} ProstAttention-Net performance. FROC analysis for detection sensitivity on CS lesions (GS~$>6$), based on 5-fold cross-validation. 
The solid blue line shows performance on the whole 24-slices volume while the dotted green line shows results considering only slices with at least one lesion as in~\cite{cao_joint_2019}. The transparent areas are 95\% confidence intervals corresponding to 2x the standard deviation.\vspace{-0.2cm}}
\includegraphics[width=0.6\linewidth]{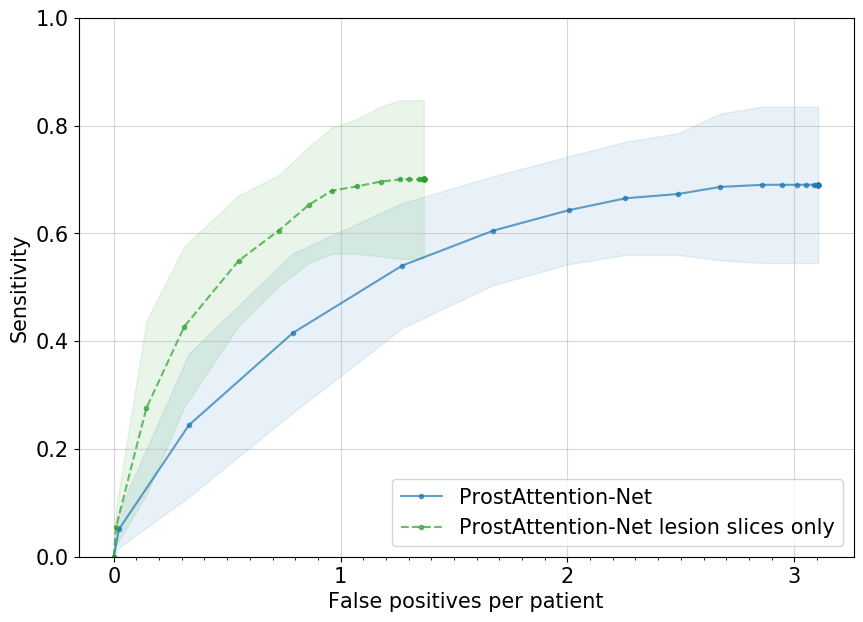}
\end{figure}

\autoref{fig:froc} shows FROC results of  ProstAttention-Net for CS (GS~$>6$) cancer detection task. The blue curve shows that our model achieves 69.0\% $\pm 14.5$\% sensitivity at 2.9 false positives per patient.
The additional green curve reported on \autoref{fig:froc} shows ProstAttention-Net performance if, instead of considering the whole transverse slices (24 slices) of each patient volume, we only perform prediction on slices with at least one annotated lesion, as in \citet{cao_joint_2019}.
This curve shows that our model achieves 68.3\% $\pm 12.1$\% sensitivity at 1 false positive per patient. Such a performance compares favourably to that of FocalNet~\citep{cao_joint_2019} trained with a crossentropy loss. 
For a 1 false positive per patient, FocalNet reaches $\sim 60$\%  with a crossentropy loss and $\sim 80$\% with their focal loss.

\subsubsection{Segmentation by Gleason score}
\label{subsubsection:results_by_gs}

Results of the FROC analysis for each GS group are presented in \autoref{tab:frocbyclass}. It reflects the model ability to both localize  lesions and assign their correct GS.
They show that detection performance for each GS group is correlated to the lesion aggressiveness: the higher the cancer GS group is, the better the detection rate gets, as also reported in \citet{cao_joint_2019}. Let us remind that this FROC by GS is pessimistic as a correctly localized lesion with an incorrect GS is both a false negative for the true class and a false positive detection for the predicted class.

\begin{figure}[tp]
\centering
\caption{\label{fig:confmatVP} Normalized confusion matrix of lesion Gleason score prediction with ProstAttention-Net. This confusion matrix is the sum of the 5 confusion matrices obtained for each validation fold. Only true positive (detected) lesions are included in this matrix. The corresponding Cohen quadratic-weighted kappa score is $0.418\pm0.138$ when results are averaged over the 5 subfolds or 0.440 when considering this total confusion matrix.}
\includegraphics[width=0.6\linewidth]{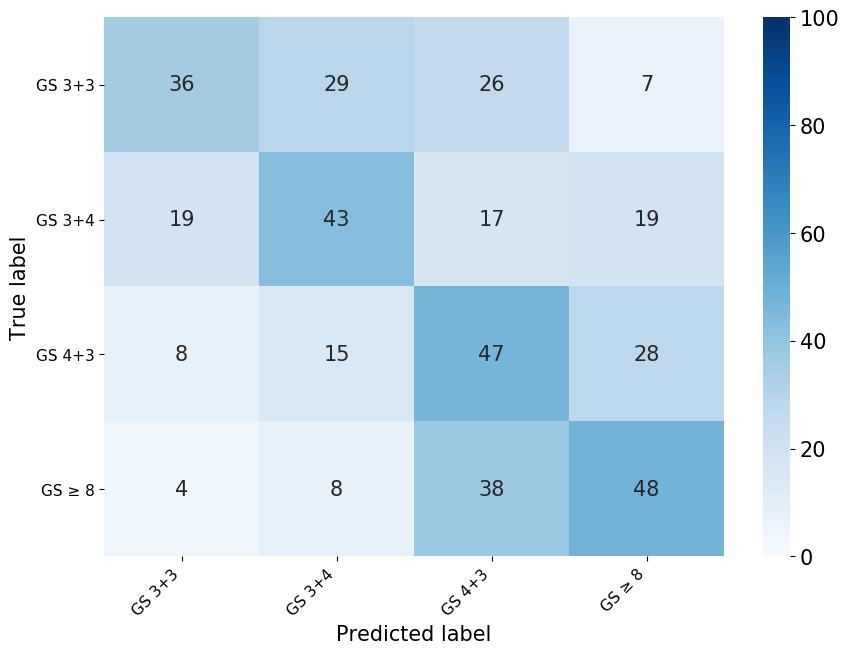}
\end{figure}

The confusion matrix in \autoref{fig:confmatVP} shows the predicted class for each detected lesion versus the reference lesion class. 
We observe that the highest percentages ($>40$\%) of detected lesions of each class are in the diagonal, thus meaning that they were assigned the correct GS. In general, few lesions that do not belong to GS~$\ge8$ class are confounded with this class (7\% of GS 6, 19\% of GS 3+4 and 28\% of GS 4+3). In the same way, few CS lesions (GS~$\ge7$) are predicted as GS 6 lesions (19\% of GS 3+4, 8\% of GS 4+3 and 4\% of GS~$\ge8$). 
Most errors correspond to GS~$\ge8$ predicted as GS 4+3 and GS 3+3 predicted GS 3+4  (respectively 38\% and 29\% of the misclassification rate for these two GS classes.) 
The corresponding quadratic-weighted Cohen kappa score is $0.418\pm0.138$ when results are averaged over the 5 subfolds and 0.440 when computed from the confusion matrix in \autoref{fig:confmatVP}.
These values are good for this challenging task and compare favourably to values reported in state-of-the-art studies: it outperforms the reported kappa in \citet{de_vente_deep_2020} for a segmentation task ($0.172 \pm 0.169$ in 5-fold cross-validation) and approaches the highest performance achieved by \citet{abraham_automated_2019} for the classification task (0.4727 with 3-fold cross-validation), both on PROSTATEx-2 train set.

\begin{figure*}[htbp]
\centering
\caption{\label{fig:predictions}\textit{Raw label maps} predictions for several images from the validation set. The images from the first and fourth columns come from a GE 3T scanner, images from the second and fifth columns from a Siemens 1.5T and the other ones come from a Philips 3T scanner. The first three columns illustrate success cases while the three last columns are failure examples for our model.}
\includegraphics[trim=40 11 30 130, clip,width=\textwidth]{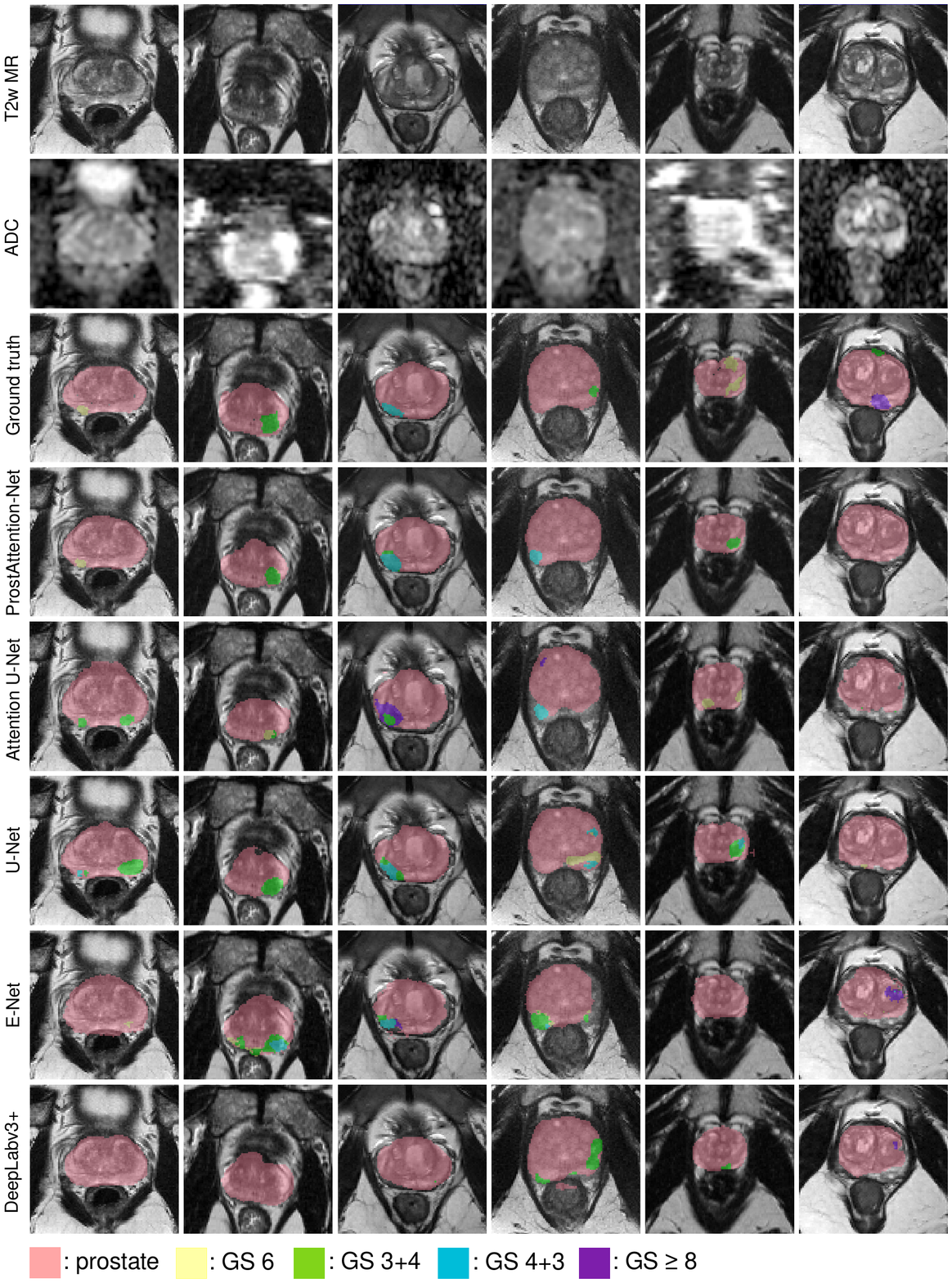}
\end{figure*}

\autoref{fig:predictions} shows prediction maps extracted from the \textit{3D raw labeled maps} outputted by ProstAttention-Net in comparison to the ground truth. The three first columns show success cases for our model while the three last columns illustrate failure cases.
In the first case (GE acquisition), a GS 6 lesion is present on this patient slice. ProstAttention-Net not only classifies this region as a lesion with the correct GS of 6 but also correctly outlines it (Dice score of 0.75). This successful detection is all the more interesting as neither of the two radiologists had identified this small and low GS lesion.
In the second case (Siemens acquisition), ProstAttention-Net correctly identifies the GS 3+4 lesion. The delineation is also good with a reported Dice score of 0.62.
On the third example (Philips acquisition), ProstAttention-Net predicts the GS 4+3 lesion with the correct grade. 
It can be noticed that a small part of the lesion is mistakenly assigned to GS 3+4, which is an interesting result since GS 4+3 is composed of a majority of grade 4 cells but also some grade 3 cells while GS 3+4 has a majority of grade 3 cells but also some grade 4 cells.
In the fourth case (GE acquisition), the GS 3+4 lesion is missed while a false positive GS 4+3 lesion is predicted on the contralateral lobe. 
Note that for this patient, the two radiologists correctly detected the lesion but also reported additional suspicious areas corresponding to false positive detections.
The fifth column (Siemens acquisition) shows a patient with two GS 6 lesions in the PZ. ProstAttention-Net succeeds in detecting one of the two lesions but assigns it the overhead GS (GS 3+4). This kind of lesion detection with a wrong GS might still help the radiologist to target biopsy on this suspicious zone. 
Finally, on the last example (Philips acquisition), both GS~3+4 and GS~$\ge8$ lesions located in the TZ and PZ areas are missed by the model while the radiologists could not identify the GS~3+4 TZ lesion but detected the GS~$\ge8$ lesion.

\subsubsection{Performance by prostate zone}
\label{subsubsection:perfbyprostatezone}
\begin{figure}[htbp]
\centering
\caption{\label{fig:frocbyzone} Performance by prostate zone. FROC analysis for detection sensitivity on CS lesions (GS~$>6$) depending on the considered prostate zone (PZ or TZ), based on 5-fold cross-validation. 
The transparent areas are 95\% confidence intervals corresponding to 2x the standard deviation.}
\includegraphics[width=0.6\linewidth]{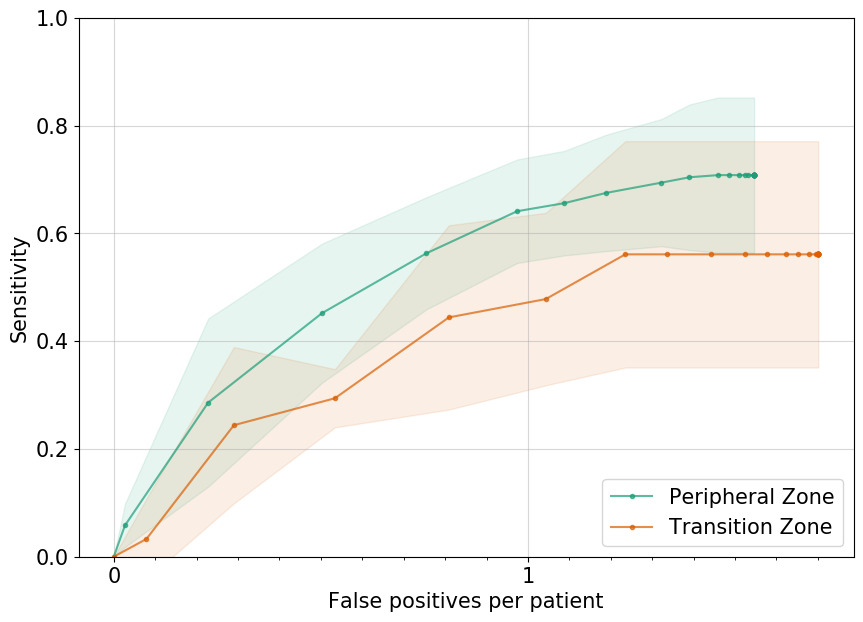}
\end{figure} 

\autoref{fig:frocbyzone} shows ProstAttention-Net performance when zonal masks were applied as described in Section \ref{subsection:base_expe_prostatttention}. As expected, performance is higher when we consider only the PZ instead of the full prostate : it reaches 70.8\% $\pm 14.4$\% sensitivity at 1.5 FP per patients. When considering the TZ only, ProstAttention-Net performance drops to 56.1\% $\pm 21.0$\% sensitivity at the same FP rate and exhibits more variation between folds. This difference is most likely due to the very low number of TZ lesions (49) in our dataset compared to the 289 PZ lesions, as reported in \autoref{tab:lesionsclass}. 
In addition, TZ lesions are difficult to distinguish from benign hyperplasia nodules \citep{oto_prostate_2010} and are significantly less detected by radiologists than PZ lesions \citep{bratan_influence_2013}.

\subsubsection{Comparison against our previous study}
\label{subsubsection:comparisonMIDL}
\begin{figure}[htbp]
\centering
\caption{\label{fig:attentionzone} Impact of the attention zone. FROC analysis for detection sensitivity on CS lesions (GS~$>6$) of the peripheral zone (PZ), based on 5-fold cross-validation. ProstAttention-Net was trained either with attention on the whole prostate as in \autoref{fig:froc} (solid green line) or with  attention on the PZ only (yellow dotted line). A mask was applied on the PZ to omit TZ lesions from the performance analysis.
The transparent areas are 95\% confidence intervals corresponding to 2x the standard deviation.
}
\includegraphics[width=0.6\linewidth]{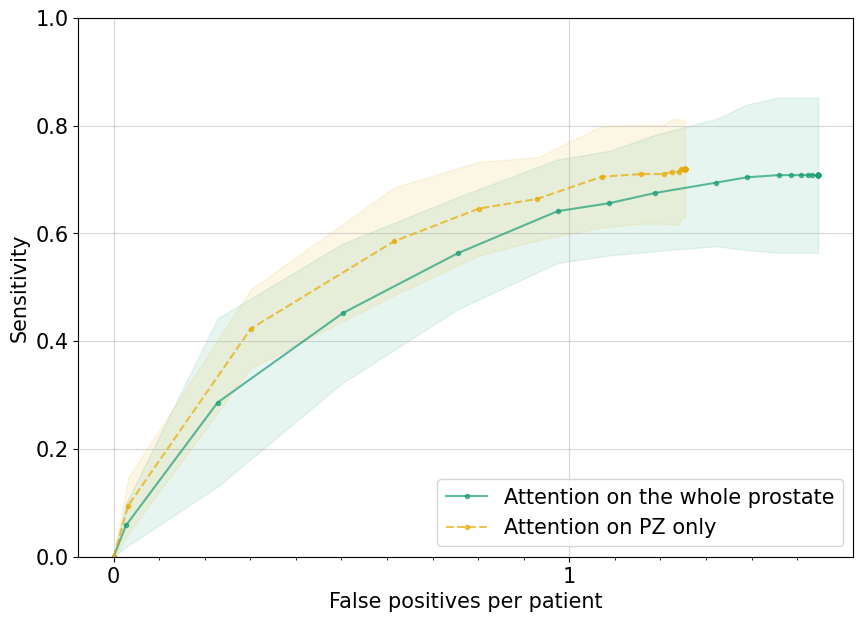}
\end{figure}

The overall CS lesion detection sensitivity reported in our preliminary study \citet{duran2020prostate} was 75.8\% $\pm 3.4$\% at 2.5 FP per patient, which is slightly higher than the value reported in this paper. As described in the Introduction, both studies present some differences.\\ First, PZ and TZ lesions were included in this study while only the prevailing PZ lesions were considered in \citet{duran2020prostate}. As seen in Section~\ref{subsubsection:perfbyprostatezone}, TZ lesions detection is not as successful as PZ lesions ones because it is a more challenging task which would benefit from an increased TZ lesion training dataset.
The addition of TZ lesions also induces a change of the attention zone, which now encompasses the whole prostate instead of the PZ region in \citet{duran2020prostate}. Impact of the attention zone was further studied to refine the comparison with results reported in \cite{duran2020prostate}. \autoref{fig:attentionzone} presents FROC analysis for CS lesions detection in the PZ of ProstAttention-Net trained either with attention on the whole prostate (green curve) - thus corresponding to the results reported in Section \ref{subsubsection:results_cslesions} - or with attention on the PZ only (yellow curve), as implemented in \cite{duran2020prostate}. As in Section \ref{subsubsection:perfbyprostatezone}, a mask was applied on the peripheral prostate zone to omit TZ lesions from the performance analysis. We observe that the sensitivity achieved with the PZ-attention model (yellow curve) is higher than that achieved with the model focusing attention on the whole prostate (green curve) with 68.4\%$\pm 8.4$\% sensitivity versus 64.6\%$\pm 9.6$\% at 1 FP. However, the difference is not significant and we can not conclude that the attention on the whole prostate degrades the performance on the PZ. \\
Secondly, the dataset is different and more heterogeneous in this study with 95 new patients from the GE and Siemens scanners, as well as 26 patients from a new Philips Ingenia scanner. In \ref{appendix:multisource}, we show that performance for this specific scanner is below to that of GE and Siemens.
The 95 new patients acquired on GE and Siemens might also contain more difficult cases. To validate this hypothesis, we tested our model on the 98 patients used in \citet{duran2020prostate}. To prevent testing on patients data used during the training phase, the same subfold split was used and the 121 new patients were added to this base. Results in \autoref{fig:frocMIDL} of \ref{appendix:perfsMIDL} show that this model performs better on the database included in \citet{duran2020prostate}, thus demonstrating the positive impact of the additional training data.\\

\subsection{Comparison with state-of-the-art segmentation architectures}

\begin{figure*}[h!]
\centering
\caption{\label{fig:frocmodels} Performance comparison of the different segmentation networks for the binary CS PCa segmentation task. FROC analysis for detection sensitivity on CS lesions (GS~$>6$), based on 5-fold cross-validation. 
The transparent areas are 95\% confidence intervals corresponding to 2x the standard deviation.}
 \includegraphics[width=0.6\linewidth]{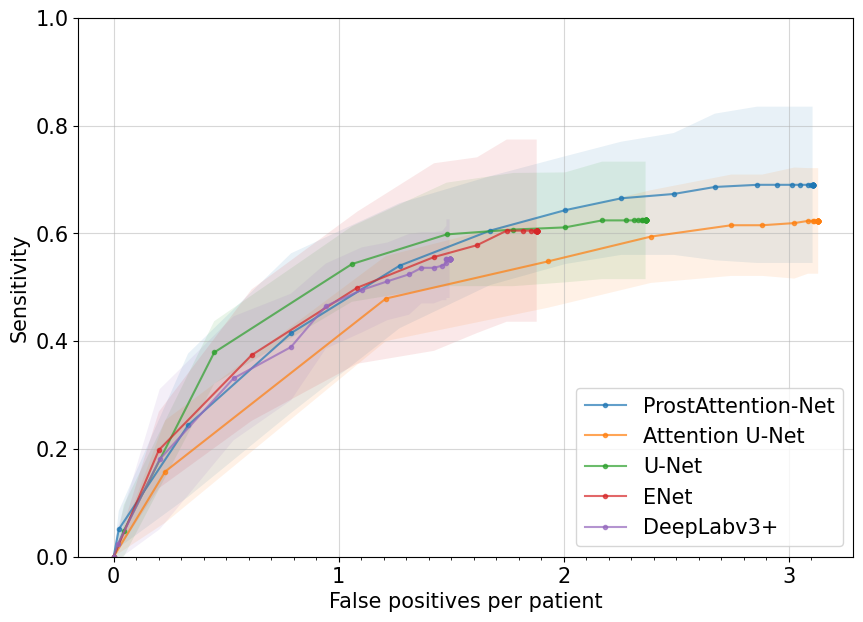}
\end{figure*}

FROC curves comparing performance of the different models for the binary CS detection task are reported on \autoref{fig:frocmodels}. In the [0, 1.5] FP interval, FROC curves of the different models mostly overlap and show similar performance. These curves are not bounded on the x-axis so that they can reach different maximal FP rates. ProstAttention-Net is shown to achieve the highest sensitivity of 69\% for a reasonable 3 FP rate per patient. 
Our model also improves the Gleason score prediction, as shown by results of the FROC analysis for each GS group reported in \autoref{tab:frocbyclass}. 
ProstAttention-Net achieves higher sensitivity at different FP rates than the other models except for the GS 6 class where ENet performs better (24\% versus 18\% sensitivity at 1.5FP).
According to a Wilcoxon one-sided signed-rank test, ProstAttention-Net performances at 1 FP and 1.5 FP are both significantly greater than U-Net and DeepLab v3+ with a p-value $<0.05$ for each of the 4 tests. For E-Net, the difference is barely not significant with a p-value of 0.051 at 1FP and 0.061 at 1.5FP. Finally, the difference with Attention U-Net is not significant either with a p-value of 0.096 at 1FP and 0.058 at 1.5FP.

\begin{table*}[!t]
\caption{\label{tab:frocbyclass}
Mean detection sensitivity per GS group in cross-validation at 1 and 1.5 FP rate per patient. Significant difference (p-value $<$ 0.05) between ProstAttention-Net and the other models according to Wilcoxon one-sided signed-rank test are shown with asterisks.} 
\centering
\resizebox{\columnwidth}{!}{
\begin{tabular}{lcccccccc}
\hline
     & \multicolumn{2}{c}{GS~$\ge8$} &  \multicolumn{2}{c}{GS 4+3} & \multicolumn{2}{c}{GS 3+4} & \multicolumn{2}{c}{GS 3+3} \\
    \cline{2-9}
    {}   & 1.5 FP & 1 FP & 1.5 FP & 1 FP & 1.5 FP & 1 FP & 1.5 FP & 1 FP\\
    \hline
    DeepLabv3+* & 0.42 & 0.42 & 0.43 & 0.43 & 0.27 & 0.26 & 0.08 & 0.08\\
    ENet & 0.48 & 0.48 & 0.39 & 0.39 & 0.39 & 0.29 & \textbf{0.24} & \textbf{0.15}\\
    U-Net* & 0.61 & 0.61 & 0.48 & 0.46 & 0.30 & 0.28 & 0.07 & 0.07\\
    Attention U-Net & 0.57 & 0.55 & 0.43 & 0.39 & 0.36 & 0.32 & 0.14 & 0.14\\
    ProstAttention-Net  & \textbf{0.74} & \textbf{0.70}& \textbf{0.61} & \textbf{0.52} & \textbf{0.40}& \textbf{0.36} & 0.18 &0.11 \\ \hline
\end{tabular}
}
\end{table*}

Cohen's quadratic weighted kappa coefficient reported in \autoref{tab:cohenkappa} also reflects the capacity of the models to correctly classify the different GS groups. We obtain a kappa score of $0.418 \pm 0.138$ with ProstAttention-Net while the corresponding kappa coefficients for DeepLabv3+, U-Net and Attention U-Net are $0.318\pm 0.114$, $0.323\pm0.157$ and $0.345\pm0.158$ respectively. ENet kappa score is shown to be very close to ProstAttention-Net score, with a mean value of $0.414\pm0.153$. 
Based on both metrics derived from the FROC analysis in \autoref{tab:frocbyclass} and the confusion matrix in \autoref{tab:cohenkappa}, we can conclude that ProstAttention-Net performs better than the state-of-the-art segmentation methods considered in this study.
This reflects the significant contribution of the proposed  attention mechanism in segmenting the different lesion classes. 
These quantitative results are confirmed by the qualitative analysis in \autoref{fig:predictions}. 
Roughly, U-Net predictions are the closest to that of ProstAttention-Net, but with a failure on the first example. Attention U-Net could identify most lesions but often with an incorrect Gleason score (first, second and third examples). However, it is the only model that could characterise the GS~6 lesion on the fourth example, but still missing the second GS~6 lesion. E-Net predictions look coarser (see the wide predicted lesion on the second example, and the poor prostate segmentation on the fourth and sixth examples). DeepLabv3+ is more conservative with fewer correctly predicted lesions but also less false positives, as reflected in the binary FROC curve (see \autoref{fig:frocmodels}).
Interestingly, on the fourth column, the other models (in particular U-Net and DeepLabv3+) did better than ProstAttention-Net.
In order to compare with performance of the state-of-the-art nnU-Net model \citep{isensee2021nnu}, we executed the nnU-Net v1.6.5 pipeline (\url{https://github.com/MIC-DKFZ/nnUNet}) in both 2D and 3D modes with full resolution versions and  default parameter setting. Input data preprocessing included all steps described in section \ref{subsubsection:data_preprocessing} except intensity scaling which is included in nnU-Net. We obtained lower performance than the models considered in our study, including U-Net, with a maximum sensitivity of $0.393\pm0.033$ in 2D and $0.389\pm0.044$ in 3D and corresponding kappa scores of $0.298 \pm 0.108$ and $0.296 \pm 0.251$, respectively. Such results might be explained by differences in the loss function (weighted versus non-weighted dice loss) as well as different hyperparameters tuning strategies (extensive grid search in our case versus fix configurations for nnU-Net).

\begin{table}[!t]
\caption{\label{tab:cohenkappa} Mean quadratic weighted Cohen's kappa score for the different models. According to Wilcoxon one-sided signed-rank test, ProstAttention-Net kappa score is not significantly greater than for the other models.}
\centering
{
\begin{tabular}{lc}
\hline
DeepLabv3+ &$0.318\pm 0.114$\\
ENet & $0.414\pm0.153$ \\ 
U-Net &  $0.323\pm0.157$ \\
Attention U-Net & $0.345\pm0.158$\\
ProstAttention-Net & $\mathbf{0.418\pm0.138}$\\
\hline
\end{tabular}}
\end{table}

\label{subsection:robustnesstomultisource}

\subsection{Performance evaluation on the PROSTATEx-2 public dataset}

A weighted kappa score of $0.120\pm0.092$ (with 1k iterations bootstrapping) was computed from the best performing subfold model on the train set of the PROSTATEx-2 dataset.
This kappa score is smaller than the reported value of  $0.172\pm0.169$ (with 1k iterations bootstrapping) achieved by \citet{de_vente_deep_2020} on PROSTATEx-2 train set but also less prone to variation thus more robust.
Let us remind that the kappa score reported by \citet{de_vente_deep_2020} was achieved with a model trained on PROSTATEx-2 lesions masks that were manually drawn with a in-house software. Unfortunately, these manually-outlined maps were not rendered public thus preventing us from fairly comparing these scores.
Our kappa score can also be compared to the kappa value of $0.13\pm0.27$ reported by the same team on the unseen PROSTATEx-2 test set. Unfortunately, we could not evaluate our model on PROSTATEx-2 test data because the challenge has closed. 

\begin{figure}[tp]
\centering
\caption{\label{fig:pred_px2}ProstAttention-Net prediction for several PROSTATEx-2 images. The ground truth consists in the lesion's center coordinates.} 
\includegraphics[trim=0 390 65 0, clip, width=0.8\linewidth]{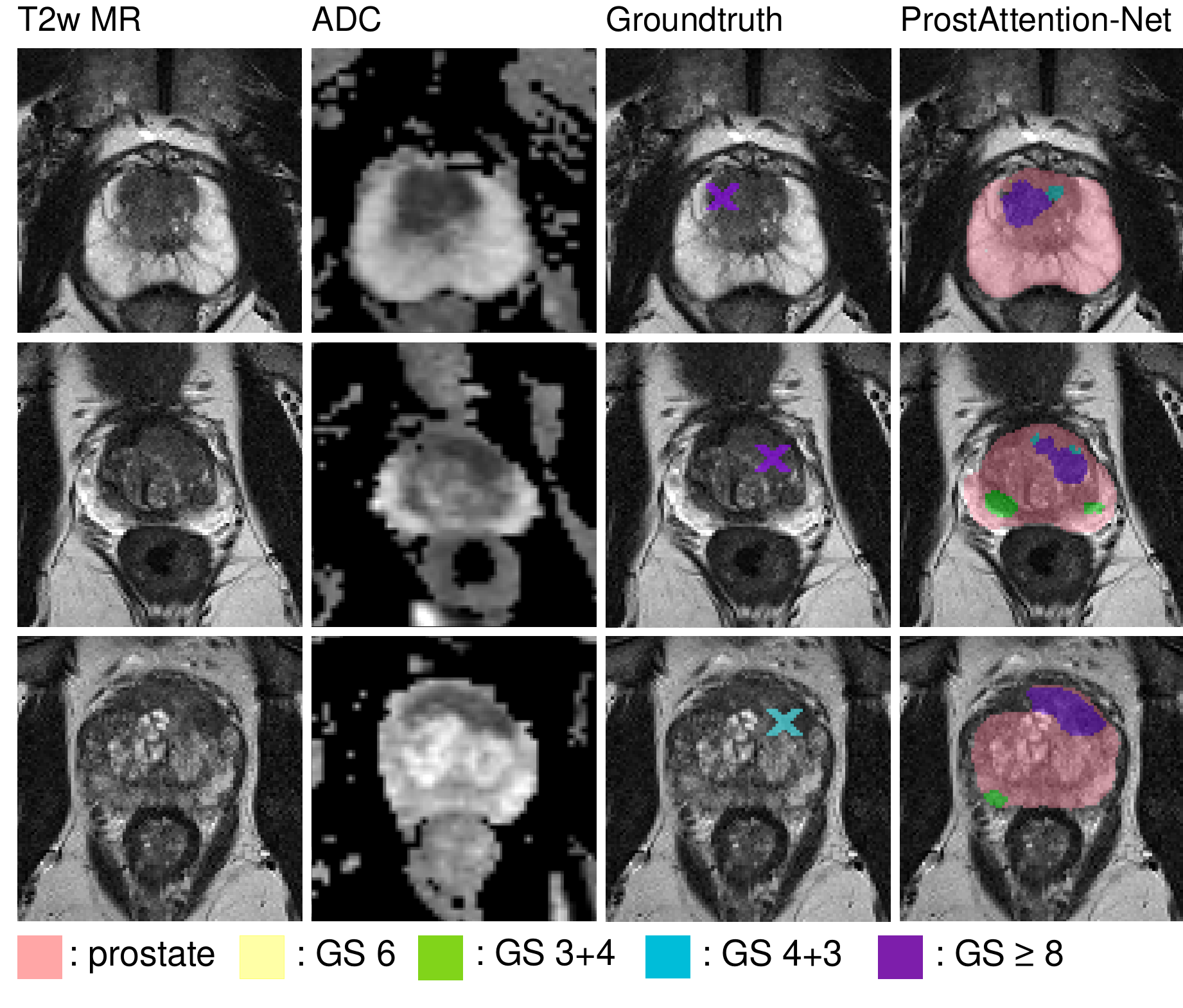} 
\end{figure}

\autoref{fig:pred_px2} illustrates ProstAttention-Net predictions on three examples of PROSTATEx-2 training images with TZ lesions. Visually, results seem satisfying: the prostate is well predicted and the TZ lesions segmentation looks correct. On the second and third examples, we can however notice small false positive lesions of GS 3+4. In addition, on the third example, the lesion is misclassified as GS~$\ge8$ instead of GS 4+3. 

Performance achieved on the PROSTATEx-2 dataset is not as good as on our private dataset.
The confusion matrix (data not shown) that serves as the basis to compute the kappa coefficient indicates that the majority of PROSTATEx-2 lesions were confounded with GS~$\ge8$. This may suggest a domain adaptation issue, even with our multi-source model.
Imaging parameters used for PROSTATEx-2 and our datasets are indeed heterogeneous and the network might need to be fine-tuned with PROSTATEx-2 data to \textit{calibrate} itself with the intensity/texture characteristics of this external dataset. Cancer lesions seem to be generally more contrasted on PROSTATEx-2 than on our dataset's ADC maps.
In addition, almost half of the lesions of PROSTATEx-2 are in the TZ while it represents less than 15\% of the lesions (49/338) in our private dataset, which is likely to impact the performance as reported in section \ref{subsubsection:perfbyprostatezone}.

\section{Discussion}

Results reveal that our proposed ProstAttention-Net can successfully segment both the prostate gland (with an average Dice on validation subfolds of $0.875\pm0.013$) and PCa lesions with performance on par with state of the art. We also demonstrate good performance in the characterisation of Gleason score grading of the detected lesions.

We showed that the addition of the zonal attention significantly improved the Gleason score grading, based on lesion-wise Cohen's kappa and FROC by GS metrics. This is concordant with recent studies showing the contribution of attention mechanism and inclusion of zonal information in PCa segmentation networks \citep{saha_end-to-end_2021, de_vente_deep_2020}.
The correct generalization of our model is surely partly due to the inclusion of multi-source data in the training. This is somehow expected and consistent with recent studies \citep{rundo_use-net_2019}. Performance on the PROSTATEx-2 dataset might be improved with the inclusion of additional TZ lesions in the training. Indeed, our private dataset only contains $\sim$14\% of TZ lesions (49/338) when PROSTATEx-2 TZ lesions represent more than 55\% of the train set lesions.

As far as we know, only two recent studies focused on PCa grading, either focusing on PCa lesion detection \citep{cao_joint_2019} or segmentation \citep{de_vente_deep_2020} tasks by GS. These studies are presented in Section~\ref{section:relatedworks}.

In order to compare to performance reported in \citet{cao_joint_2019} for the binary CS lesion detection task with the proposed FocalNet model, we computed FROC on lesion slices only, as performed in their work (see Section~\ref{subsubsection:results_cslesions}).
In comparison to FocalNet, the sensitivity achieved by ProstAttention-Net at 1 FP is 10\% higher than FocalNet trained with crossentropy loss but 15\% lower than the sensitivity reached with the combination of their focal and mutual finding losses.
We evaluated the impact of the focal loss (FL) in the training of ProstAttention-Net, by considering two different scenarios. First, we used the classical focal loss only \citep{lin_focal_2017} in the lesion decoding branch of ProstAttention-Net, instead of the combination of weighted cross-entropy and Dice losses in \autoref{eq:losslesion}. Second, the standard focal loss was combined to the weighted Dice loss. Performance in both cases did not improved compared to the original loss (data not shown), when replacing CE by the FL losses, unlike results reported by \cite{cao_joint_2019}. One explanation is that we used weighted versions of CE which already account for the class imbalance problem unlike the standard CE loss used by \citet{cao_joint_2019}. Replacing this weighted CE by FL thus did not allow any significant performance gain.
Another difference is that we did not include the focal weight \textit{q} described in their adapted focal loss for ordinal encoding. 

The lower performance for CS PCa detection might also be explained by the smaller number of lesions in our dataset (2:1 ratio). Also, FocalNet was trained and tested on a more homogeneous dataset including 417 patients all acquired on Siemens 3T scanners, with the same imaging parameters. 
The authors do not report performance on the PROSTATEx-2 challenge dataset so that a fair comparison is not possible.

Regarding comparison for the GS grading task, comparison with the study of \cite{cao_joint_2019} is not straightforward.
From our understanding, the FROC metrics reported in \citet{cao_joint_2019} only focus on the binary detection of the lesions. Indeed, the FROC by GS expresses the quantity of detected lesions of a given GS, without consideration of the Gleason score assigned by the model. This analysis is different from ours where a true positive is considered so when it overlaps a true lesion and it is assigned the correct Gleason score; otherwise the detected lesion is counted as a false negative for this class and a false positive for the predicted class.
Applying the same rules than \cite{cao_joint_2019} to compute FROC by GS, we get sensitivities of 0.91, 0.81, 0.61 and 0.40 respectively for GS~$\ge8$, GS 4+3, GS 3+4 and GS 6 groups (instead of the 0.74, 0.61, 0.40, 0.18 values reported in \autoref{tab:frocbyclass} at 1.5 FP per patient).
\citet{cao_joint_2019} also do not use confusion matrix or Cohen's kappa metrics to report GS prediction performance, yielding to a difficult comparison. The only evaluation of GS prediction consists in the four binary-classifications tasks presented in Section \ref{section:relatedworks}.

Also, our model does not only detect points of interest as in \citet{cao_joint_2019} but accurately segments the lesions, as well as the whole prostate. 
We consider that these two segmentation outputs (shape of the lesions and of the prostate) are of great clinical interest. The prostate segmentation, as an example, allows the clinicians to visualize where the network did actually search for lesions, which can be useful to assess confidence in the CAD model.
Our will to achieve accurate segmentation requires to define more constraining rules to define true positive detection, indeed based on the intersection between the ground truth and the prediction and not on a distance between centroids as done in \cite{cao_joint_2019}. This may be at the cost of a slight performance loss when evaluating detection tasks.

In order to compare with \citet{de_vente_deep_2020}, we computed lesion-wise Cohen's kappa score.
Including false negative lesions in the GS 6 class as they proposed, we achieved a kappa score of $0.511 \pm 0.076$ for ProstAttention-Net (see ~\ref{appendix:confmatGS6FN}) which is much higher than their reported value of $0.172 \pm 0.169$ for their 5-fold cross-validation model. When we test the generalization capacities of ProstAttention-Net on PROSTATEx-2 training dataset (unseen dataset), we get a lesion-wise kappa score of $0.12 \pm 0.09$ (with 1000 iterations bootstrapping), which is close to the reported value of $0.13\pm0.27$ in \citet{de_vente_deep_2020} on PROSTATEx-2 test set and less prone to variation. 
Note that \citet{de_vente_deep_2020} trained their model on PROSTATEx-2 training dataset (with added manually contoured lesions), so that the reported performance on the test set above does not correspond to generalization performance achieved on unseen data from a new domain, as for ProstAttention-Net. 

In addition, while \citet{de_vente_deep_2020} only consider CS lesions (GS $\ge$ 7), our model also performs the segmentation of the GS 6 lesions, which is a crucial grade to include because of the lower sensitivity of radiologists for this grade (0.480 and 0.589 respectively for the two radiologists on our dataset - data not shown) and with a high clinical interest for active surveillance. 

Comparison with these two reference studies indicates that the GS performances achieved by ProstAttention-Net for the segmentation and GS grading of PCa can be considered as on par with the top-performing methods.

Results reported in this study as well as comparison with state of the art underline the strong impact of the chosen metric for performance evaluation of multiclass detection and segmentation models. This topic is currently actively discussed in the community \citep{reinke_common_2021}. In this paper, we put emphasis on the weighted Cohen's kappa metric to enable comparison with state-of-the-art models, but we also reported other metrics derived from the FROC analysis. The kappa metric should indeed be considered with care since it is computed from the confusion matrix that only contains positive detections. As a consequence, a model that detects very few lesions (poor sensitivity) but classifies them with the correct Gleason score can render a high kappa value. A misclassification (lesion identified as GS~3+4 instead of GS~$\ge$8 for example) is more penalized than a missed lesion. Each detected lesion has a greater influence in the confusion matrix than in the FROC analysis, which may induce the observed high standard deviation in the Cohen's kappa score in a cross-validation experiment but also between cross-validation replicates. The sensitivity by class derived from the FROC curves (cf. \autoref{tab:frocbyclass}) seems more robust and less prone to variations. Although this metric is very severe, we consider it more meaningful than the kappa coefficient, as it translates both the percentage of identified lesions and their characterization. This metric is not perfect either, as a correctly detected lesion  with the wrong GS will be considered as an error for the ground truth class and a FP for the predicted class. A metric that would weight the error, such as kappa score, but also consider sensitivity is still missing.

In order to enhance the performance, we plan to include additional MRI exams in the training database in conjunction with a domain adaptation method to harmonize the data. A weakly-supervised approach will also be considered to allow the inclusion of clinical data with coarse annotations as in the PROSTATEx-2 dataset. The contribution of additional modalities (high b-value diffusion MR, perfusion) should also be studied.
Furthermore, it could be valuable to use an ordinal encoding as in \citet{cao_joint_2019} and \citet{de_vente_deep_2020}, to fully exploit the existing hierarchy between each GS group.

\section{Conclusion}
Our ProstAttention-Net model outperforms well-tuned U-Net, Attention U-Net, ENet and DeepLabv3+ at the task of detecting clinically significant (GS~$>6$) prostate cancers. 
It achieves 69.0\% $\pm 14.5$\% sensitivity at 2.9 false positives per patient on the whole prostate and 70.8\% $\pm 14.4$\% sensitivity at 1.5 false positives when considering the PZ only, where the vast majority of lesions are localized. Regarding the automatic GS group grading, Cohen's quadratic weighted kappa coefficient is $0.418 \pm 0.138$, which is the best reported lesion-wise kappa for GS segmentation to our knowledge.
The attention zone is efficient when it consists in the whole prostate or the PZ only as in \citet{duran2020prostate}. 
It succeeds in learning from an heterogeneous database containing T2 and ADC exams acquired on scanners from three different manufacturers, with different static magnetic fields (1.5T and 3T), sequence and reconstruction parameters. Performance that we report in this study is a good indicator that our model is robust to the heterogeneity of the training database. 
This is the first time a deep convolutional network trained on an heterogeneous dataset can successfully segment prostate lesions as well as predict their Gleason score. 
Our model has encouraging generalization capacity on the PROSTATEx-2 dataset ($\kappa=0.120\pm0.092$), which is one important CAD limitation nowadays. 
We provided elements of comparison with state-of-the art methods, however a fair comparison with some previous works is difficult as the datasets, outputs and evaluation metrics are not the same and the perfect metric is still missing. 
\section*{CRediT authorship contribution statement}
\textbf{Audrey Duran:} Conceptualization, Data curation, Methodology, Software, Validation, Investigation, Writing - Original Draft, Visualization. \textbf{Gaspard Dussert:} Software, Validation. \textbf{Olivier Rouvière:} Resources : clinical imaging data acquisition and curation. \textbf{Tristan Jaouen:} Data curation. \textbf{Pierre-Marc Jodoin:} Conceptualization, Methodology, Software, Writing - Review \& Editing. \textbf{Carole Lartizien:} Conceptualization, Methodology, Validation, Writing -  Original Draft, Review \& Editing, Supervision, Funding acquisition.

\section*{Acknowledgments}
This work was supported by the RHU PERFUSE (ANR-17-RHUS-0006) of Université Claude Bernard Lyon 1 (UCBL), within the program “Investissements d'Avenir” operated by the French National Research Agency (ANR).

\bibliography{refs}
\newpage
\section*{Appendices}
\appendix
\section{Lesion distribution by Gleason score, scanner type and prostate zone}
\setcounter{table}{0}    
\label{appendix:lesionbyclass}
The following Appendix describes the lesion distribution for the 219 patients of our private dataset, corresponding to 67 Siemens Symphony, 126 GE Discovery and 26 Philips Ingenia acquisitions.
\begin{table*}[h] 
\caption{\label{tab:lesionbyclass}Number of PZ and TZ lesions per Gleason score and scanner type}
\centering
\begin{tabular}{lllllll}
\hline
Scanner  & GS 6 & GS 3+4 & GS 4+3 & GS 8 & GS~$\ge9$ & Total \\
\hline
Siemens Symphony & 44 & 39 & 12 & 11 & 6  & 112 \\
Philips Ingenia & 7 & 23 & 11 & 3  & 2 & 46 \\
GE Discovery & 53 & 64 & 33 & 16 & 14 & 180 \\
\hline
\textbf{sum} & 104 & 126 & 56 & 30  & 22 & 338  \\
\hline
\end{tabular}
\end{table*}

\begin{table*}[h] 
\caption{\label{tab:PZlesionbyclass}Number of PZ lesions per Gleason score and scanner type}
\centering
\begin{tabular}{lllllll}
\hline
Scanner  & GS 6 & GS 3+4 & GS 4+3 & GS 8 & GS~$\ge9$ & Total \\
\hline
Siemens Symphony & 36 & 34 & 11 & 7 & 5 & 93 \\
Philips Ingenia & 5  & 19 & 11 & 3 & 2 & 40 \\
GE Discovery & 42 & 58 & 30 & 14  & 12 & 156  \\
\hline
\textbf{sum} & 83 & 111 & 52 & 24 & 19 & 289 \\
\hline
\end{tabular}
\end{table*}

\begin{table*}[h] 
\caption{\label{tab:TZlesionbyclass}Number of TZ lesions per Gleason score and scanner type}
\centering
\begin{tabular}{lllllll}
\hline
Scanner & GS 6 & GS 3+4 & GS 4+3 & GS 8 & GS~$\ge9$ & Total \\
\hline
Siemens Symphony & 8 & 5 & 1 & 4    & 1  & 19 \\
Philips Ingenia & 2 & 4 & 0 & 0 & 0 & 6  \\
GE Discovery & 11 & 6 & 3  & 2 & 2 & 24 \\
\hline
\textbf{sum} & 21 & 15 & 4  & 6 & 3 & 49 \\
\hline
\end{tabular}
\end{table*}

\section{ProstAttention-Net confusion matrix when considering all lesions}
\setcounter{figure}{0}    
\label{appendix:confmatGS6FN}
\begin{figure*}[h] 
\centering
\caption{\label{fig:confmatGS6FN} Normalized confusion matrix of lesion Gleason score prediction with ProstAttention-Net. In this matrix, lesions that were not predicted as CS lesions are considered as predicted GS 6.
The corresponding quadratic-weighted Cohen kappa score is $0.511\pm0.076$ when results are averaged over the 5 subfolds or 0.513 when considering this total confusion matrix.}
\includegraphics[width=0.6\linewidth]{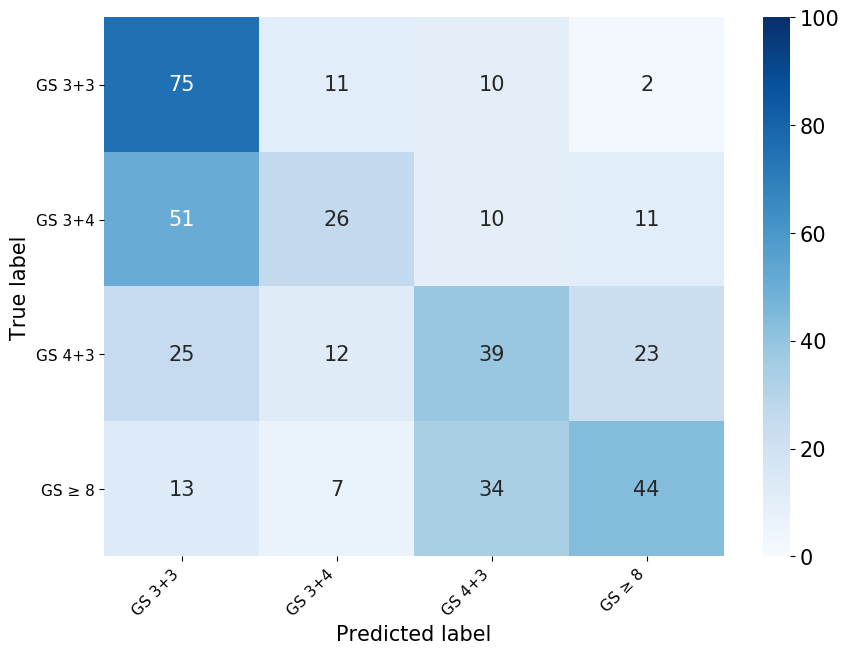} 
\end{figure*}

\section{Robustness to multi-source learning}
\label{appendix:multisource}

As reported in \autoref{tab:paramsMRI}, data of our database were acquired on three different scanners, from three manufacturers, with different magnetic field strengths and with disparate imaging parameters. In this way, each subdataset constitutes a different source and we can wonder if having a distinct model for each source would perform better than a unique model trained on a bigger but heterogeneous dataset. 
Based on the model in \autoref{fig:attunet}, two training configurations were tested:
\begin{itemize}
    \item \textbf{Multi-source learning} which consists in training the model with all 219 exams regardless of the MRI manufacturer (126 GE, 67 Siemens and 26 Philips). This configuration corresponds to the ProstAttention-Net model trained following the protocol described in Section \ref{subsection:base_expe_prostatttention}. 
    \item \textbf{Single source learning} which consists in performing three separate trainings of ProstAttention-Net, GE-only exams (126 patients), Siemens-only exams (67 patients) and Philips-only exams (26 patients), respectively. Each source-specific model was then tested on images acquired on the same scanner.  Each model was trained and validated using a 5-fold cross-validation. 
    The same patient splits were used in the multi-source and single-source experiments. In this way, a patient that was affected to one specific subfold in a single-source experiment was assigned to the same subfold in the multi-source experiments.
\end{itemize}

\begin{figure}[h]
\centering
\subfloat[]{\includegraphics[width=0.5\linewidth]{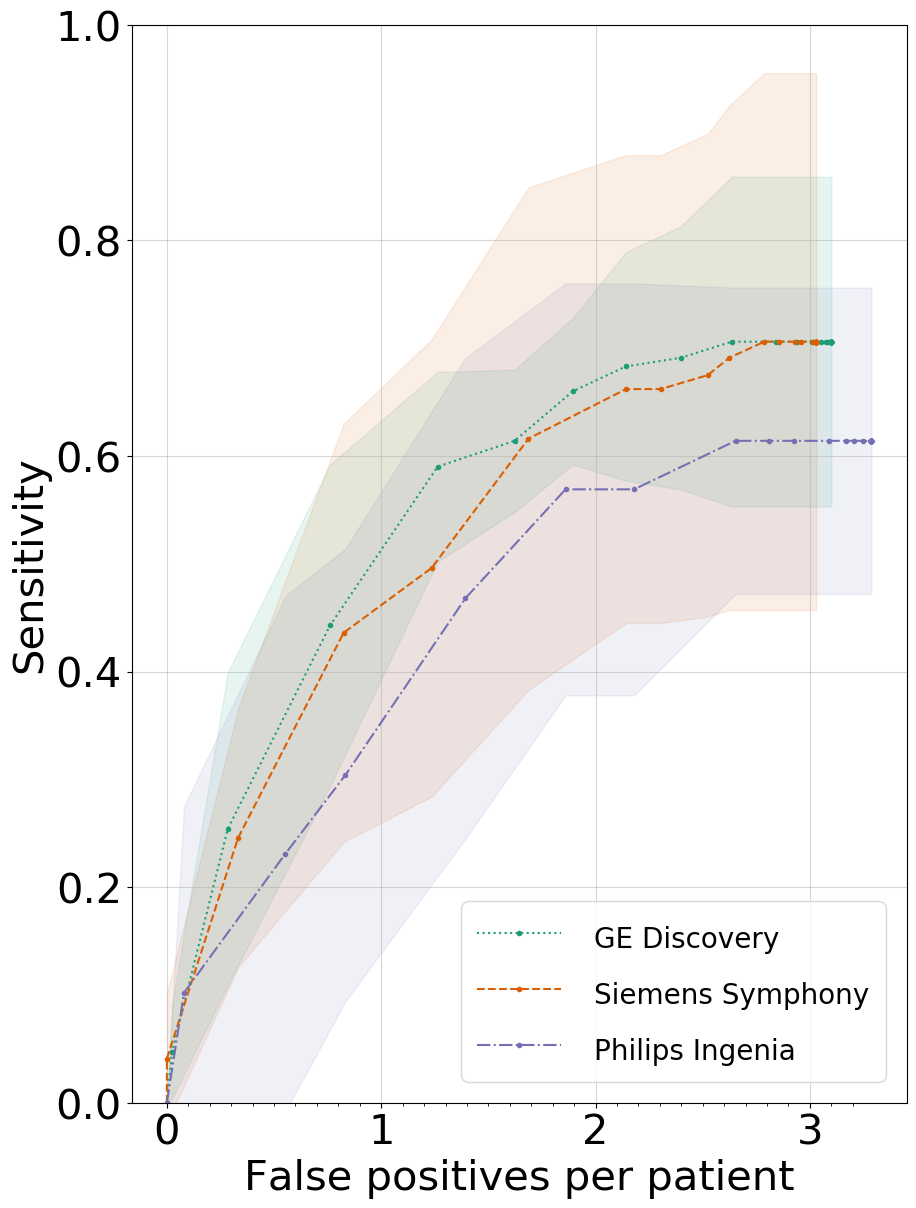}\label{fig:multisource}}
\subfloat[]{\includegraphics[width=0.5\linewidth]{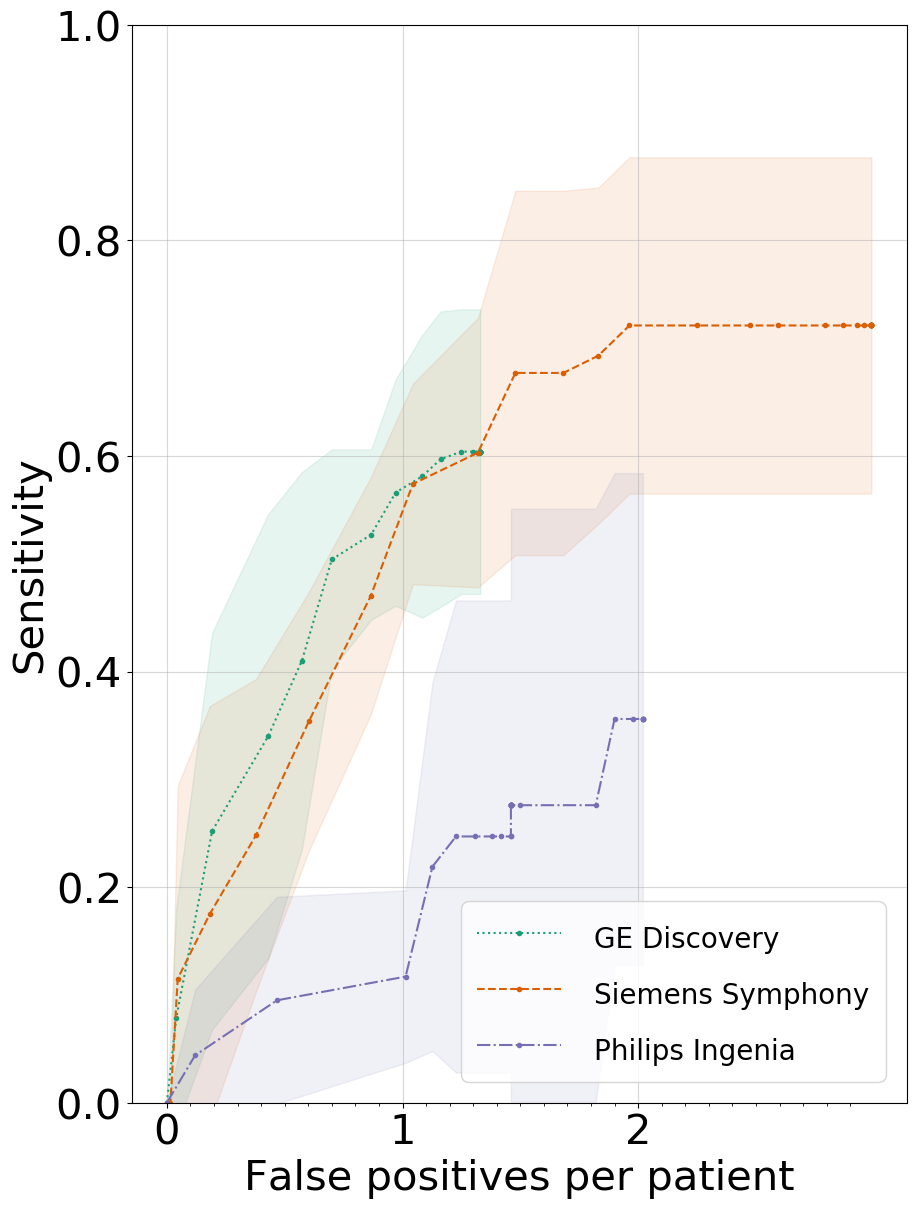}\label{fig:singlesource}}\\
\caption{Impact of multi-source learning. FROC sensitivity curves of CS lesions (GS~$>$6) with a 5-fold cross-validation when testing on each scanner independently. Results for (a) multi-source training (b) single-source training.}  
\label{fig:frocimpactsources}
\end{figure}

~\autoref{fig:frocimpactsources} shows model performance for each of the three scanners independently.
Figure \autoref{fig:multisource} reports performance achieved based on the multi-source training strategy, where ProstAttention-Net is trained with all data from the three sources corresponding to 219 exams. We can see that performance for GE and Siemens patients is similar ($\sim65-67$\% sensitivity at 2FP per patient) when Philips patients show lower performance ($\sim57$\% sensitivity at 2FP). 

Figure \autoref{fig:singlesource} reports performance of the three single-source models trained and tested on the same source. Except for Siemens, performance drops when trained on single-source only. 
This suggests that the Philips very small dataset (26 patients) and GE dataset (126 patients) benefit from the additional training data of other scanners. The loss of sensitivity observed for the Siemens patients in the multi-source learning strategy might be explained by the 2:1 ratio that exists between the number of exams acquired on 3T scanners (GE discovery and Philips Ingenia) and 1.5T scanner (Siemens Symphony) in the multi-source model : the network might indeed focus a little bit more on the pattern of 3T data (see example illustration on \autoref{fig:anapath}), at the expense of an alteration of  sensitivity to detect lesions in 1.5T Siemens exams.

\newpage
\section{Impact of the dataset}
\setcounter{figure}{0}    
\label{appendix:perfsMIDL}
\begin{figure*}[h]
\centering
\caption{\label{fig:frocMIDL} Impact of the train and test datasets on ProstAttention-Net. FROC analysis for detection sensitivity on CS lesions (GS~$>6$), based on 5-fold cross-validation.
Solid lines show performance of ProstAttention trained on the 219 patients described in \ref{appendix:lesionbyclass}, with attention on the whole prostate and tested either on the validation folds from the same 219 patients dataset (green curve) or on the validation folds encompassing the subset of 98 patients reported in \citet{duran2020prostate} (grey curve). The dotted brown curve reports performance of ProstAttention trained and tested on the same 98 subdataset, with attention on the PZ only as in \citet{duran2020prostate}. A mask was applied on the PZ to omit TZ lesions from the performance analysis.
The transparent areas are 95\% confidence intervals corresponding to 2x the standard deviation.
}
 \includegraphics[width=0.6\linewidth]{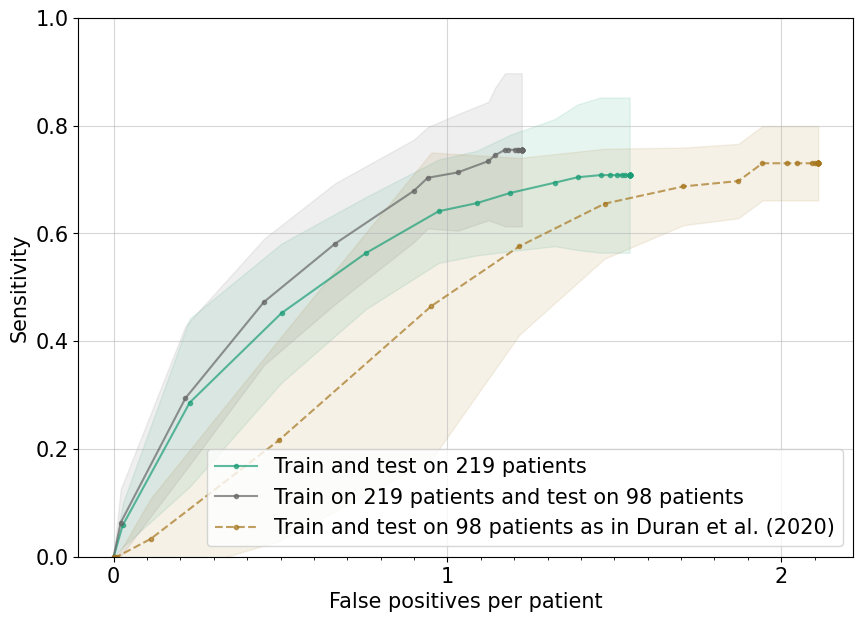}
\end{figure*}

\newpage
\section{The PROSTATEx-2 challenge dataset}
\setcounter{table}{0}    
\label{appendix:px2lesions}

\begin{table*}[h] 
\caption{\label{tab:paramsMRI_PROSTATEx-2}Imaging parameters of PROSTATEx-2 data}
\centering
\resizebox{\columnwidth}{!}{
\begin{tabular}{ccccccccc}
\hline 
Scanner & Field & Sequence & $T_R$  & $T_E$ & FOV & Matrix & Voxel dimension & Max b-value\\ &strength& &(ms)&(ms)&(mm)&(voxels)&(mm)&(s/mm$^2$)\\
 \hline
Siemens Skyra  & 3T & T2w & 5660 & 104 & $192\times 192$ & $384 \times 384$& $.5\times.5\times3$ & - \\
Siemens Skyra & 3T & ADC & 2700 & 63 & $256\times 168$ & $128 \times 84$& $2\times2\times3$ & 800\\
Siemens TrioTim & 3T & T2w & 4480 & 103 & $180\times 180$ & $320 \times 320$& $.56\times.56\times3$ & - \\
Siemens TrioTim & 3T & ADC & 2700 & 63 & $212 \times 256$& $106\times 128$ &  $2\times2\times4$ & 800\\ \hline
\end{tabular}
}
\end{table*}

\begin{table*}[h]
\caption{Lesions distribution by Gleason score in peripheral zones (PZ) and transition zones (TZ) in the PROSTATEx-2 dataset. }\label{tab:Px2lesionsclass}
\centering
\begin{tabular}{cccccc}
\hline
 & GS 3+3 & GS 3+4 & GS 4+3  & GS~$\ge8$ & Total\\ 
 \hline
PZ & 14 & 21 & 9 & 6 & 50\\
TZ & 22 & 20 & 11 & 9 & 62 \\
Total & 36 & 41 & 20 & 15 & 112 \\
\hline
\end{tabular}
\end{table*}

\end{document}